\documentclass[twocolumn,prl]{revtex4-1}
\pdfoutput=1
\usepackage[latin9]{inputenc}
\usepackage{color}
\definecolor{note_fontcolor}{rgb}{0, 0.335938, 1}
\usepackage{textcomp}
\usepackage{amsbsy}
\usepackage{amstext}
\usepackage{amssymb}
\usepackage{graphicx}
\usepackage{esint}
\usepackage[unicode=true,
 bookmarks=false,
 breaklinks=false,pdfborder={0 0 1},backref=false,colorlinks=true]
 {hyperref}
\hypersetup{pdftitle={null},
 pdfauthor={null},
 pdfsubject={null},
 unicode=false,pdftoolbar=false,pdfmenubar=false,pdffitwindow=false,pdfstartview={FitH},pdfcreator={null},pdfnewwindow=true,linkcolor=blue,citecolor=blue,filecolor=blue,urlcolor=blue}

\makeatletter

\newcommand{\lyxmathsym}[1]{\ifmmode\begingroup\def\b@ld{bold}
  \text{\ifx\math@version\b@ld\bfseries\fi#1}\endgroup\else#1\fi}

\usepackage{epstopdf}
\usepackage{color}

\usepackage{multirow}

\makeatother

\makeatletter
\def\@ssect@ltx#1#2#3#4#5#6[#7]#8{%
  \def\H@svsec{\phantomsection}%
  \@tempskipa #5\relax
  \@ifdim{\@tempskipa>\z@}{%
    \begingroup
      \interlinepenalty \@M
      #6{%
       \@ifundefined{@hangfroms@#1}{\@hang@froms}{\csname @hangfroms@#1\endcsname}%
       {\hskip#3\relax\H@svsec}{#8}%
      }%
      \@@par
    \endgroup
    \@ifundefined{#1smark}{\@gobble}{\csname #1smark\endcsname}{#7}%
  }{%
    \def\@svsechd{%
      #6{%
       \@ifundefined{@runin@tos@#1}{\@runin@tos}{\csname @runin@tos@#1\endcsname}%
       {\hskip#3\relax\H@svsec}{#8}%
      }%
      \@ifundefined{#1smark}{\@gobble}{\csname #1smark\endcsname}{#7}%
      \addcontentsline{toc}{#1}{\protect\numberline{}#8}%
    }%
  }%
  \@xsect{#5}%
}%
\makeatother

\begin{document}
\title{Two-level systems in superconducting quantum devices due to trapped quasiparticles}

\author{S. E. de Graaf$^{1\dagger}$}
\author{L. Faoro$^{2,3}$}
\author{L. B. Ioffe$^{3,4}$}
\author{S. Mahashabde$^{5}$}
\author{J. J. Burnett$^{1}$}
\author{T. Lindstr\"om$^{1}$}
\author{S. E. Kubatkin$^{5}$}
\author{A. V. Danilov$^{5}$}
\author{A. Ya. Tzalenchuk$^{1,6}$}
\affiliation{$^{1}$National physical laboratory, Hampton Road, Teddington, TW11
0LW, UK}
\affiliation{$^{2}$Sorbonne Universite, Laboratoire de Physique Theorique et Hautes Energies, UMR 7589 CNRS, Tour 13, 5eme Etage, 4 Place Jussieu, F-75252 Paris 05, France }
\affiliation{$^{3}$Department of Physics, University of Wisconsin-Madison, Madison, Wisconsin 53706, USA  }
\affiliation{$^{4}$Google Inc., Venice, CA 90291, USA}
\affiliation{$^{5}$Department of Microtechnology and Nanoscience MC2,
Chalmers University of Technology, SE-41296 Goteborg, Sweden}
\affiliation{$^{6}$Royal Holloway, University of London, Egham, TW20 0EX, UK}
\affiliation{$^{\dagger}$\textrm{sdg@npl.co.uk}}
\maketitle


\textbf{A major issue for the implementation of large scale superconducting quantum circuits  is the interaction with interfacial two-level system defects (TLS) that leads to qubit relaxation and impedes qubit operation in certain frequency ranges that also drift in time.  
Another major challenge comes from non-equilibrium quasiparticles (QPs) that result in qubit dephasing and relaxation. }
\textbf{
In this work we show that such QPs can also serve as a source of TLS. 
Using spectral and temporal mapping of TLS-induced fluctuations in frequency tunable resonators, we identify a subset of the general TLS population that are highly coherent TLS with a low reconfiguration temperature $\sim$ 300 mK, and a non-uniform density of states.
These properties can be understood if these TLS are formed by QPs trapped in shallow subgap states formed by spatial fluctutations of the superconducting order parameter $\Delta$. 
Magnetic field measurements of one such TLS reveals a link to superconductivity. Our results imply that trapped QPs can induce qubit relaxation.}

It is becoming increasingly evident that the most coherence limiting TLS reside outside the qubit junctions
 on the surface of metals and dielectrics surrounding them \cite{klimov2018,lisenfeld2019,burnett2014,degraaf2017,bilmes2019} and their slowly fluctuating dynamics poses a significant challenge for quantum computation \cite{supremacy,klimov2018,lisenfeld2019,schlor2019, burnett2019}.
A wide range of techniques has been developed to study and mitigate such TLS, by developing qubits
and resonators into probes of a wider range of material properties \cite{schneider2019,degraaf2017,degraaf2018,grabovskij2012,lisenfeld2019,geaney2019,degraaf2015, lisenfeld2015,lisenfeld2015b}.
However, charged surface TLS and paramagnetic impurities also result in a stochastic and locally varying backdrop for QPs in the superconductor itself. It is thus important to consider the implications on the ever-present excess number of non-equilibrium QPs in superconducting quantum devices \cite{Henrique2019, jin2015, devoretPRL} which must be eliminated in order to increase qubit dephasing times. It is also likely that trapped QPs
are responsible for the non-equilibrium relaxation of transmons recently observed \cite{devoretPRL}. 
Non-equlibrium QPs can be generated by electromagnetic radiation \cite{devisser2012} and from rare high energy cosmic particles impinging the sample, the latter inducing correlated errors in all qubits of a surface code architecture \cite{moore2012}. 

Here we reveal the implications of a spatially fluctuating $\Delta$ on such QPs. 
Scanning tunneling spectroscopy studies typically find spatial variations in $\Delta$ of the order of $\delta\Delta = 10-20\%$  in moderately
disordered superconductors \cite{carbillet2019,lemarie2013,Liao2019}.
$\delta\Delta$ gets smaller in very clean superconductors, however,
even in exceptionally clean films with negligible intrinsic magnetic
disorder an everpresent surface spin density
of the order of $\sim5\cdot10^{17}$ m$^{-2}$ \cite{degraaf2017, anton2013, saveskul2019} results in both flux noise \cite{degraaf2017, quintana2017, anton2013, kumar2016, paladino2014,muller2017} and spatially
non-uniform gap suppression \cite{saveskul2019}. In thin films the gap is also non-uniformely suppressed
by the Altshuler-Aronov effect  due to impurity scattering \cite{Altshurer_Aronov, carbillet2019}, as well as thickness variations \cite{ivry2014}, particularly relevant for Al \cite{aldelta}. The clean, low resistance NbN films that we chose to study in this work are characterized by a low carrier density \cite{Wong2017}. In such materials it is very likely that the DOS varies significantly, resulting in gap variations. 

\begin{figure*}[t!]
\includegraphics[scale=0.6]{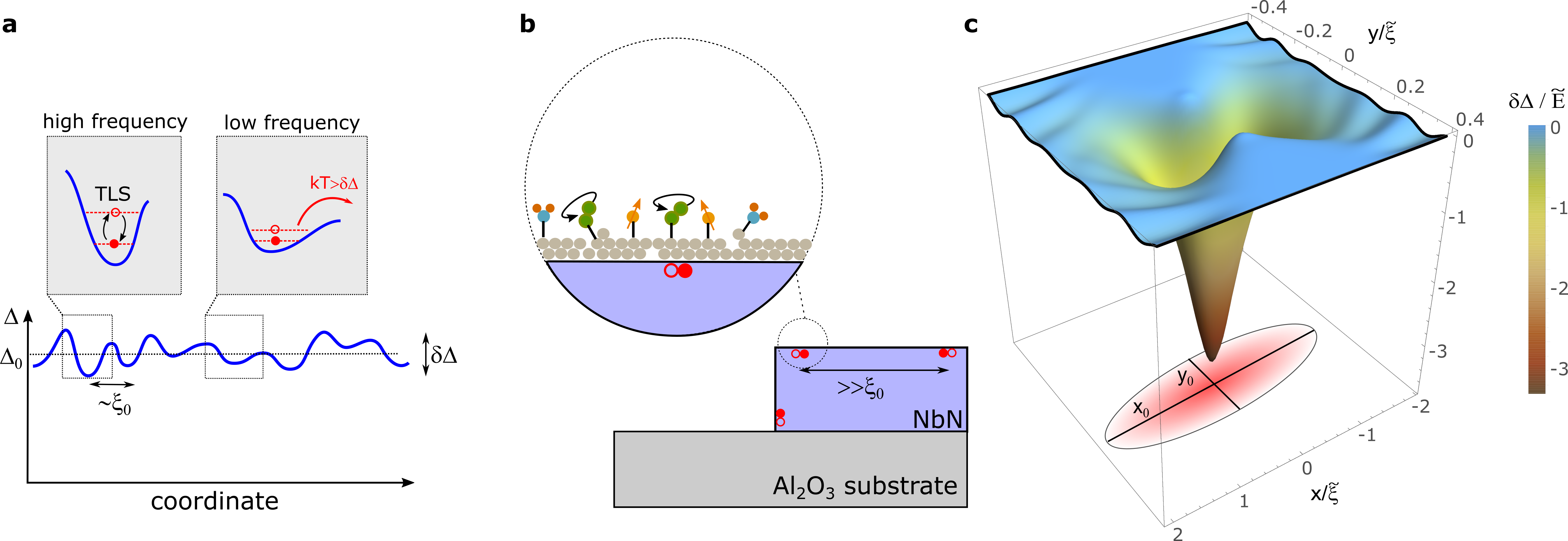}
\caption{qTLS formation. a) Illustration of typical fluctuations of the order parameter $\Delta$ and the formation of wells with multiple bound states in which quasiparticles can be trapped, effectively forming a TLS (`qTLS'). The scale of the trap determines the qTLS frequency and escape rate. b) qTLS are separated by large distances compared to length-scales at which fluctuations in $\Delta$ occur, fluctuations which may originate from impurity scattering or surface spins. Importantly the qTLS are separated in distance from the superconductor oxide surface where conventional TLS defects, adsorbants etc. reside, and form the usual bath of TLS.  c) Numerically evaluated typical elongated fluctuation of the order parameter capable of hosting multiple bound states for quasiparticles, forming a qTLS. Here $\tilde{\xi}=\sqrt{\Delta_0/E} \xi_0$, the scaling assumes $(k_F \xi_0)^{-1}= 0.14$, $\Delta_0/E = 40$, appropriate for our NbN samples.  }
\label{fig:sketch} 
\end{figure*}

Crucially, 
we find that the optimal fluctuation able to trap a QP only has a few bound states. As a result a trapped QP behaves similar to `textbook' TLS, as shown in Fig. \ref{fig:sketch}. We hereafter refer to this type of quasiparticle TLS as `qTLS', which coexist with a large number of conventional glassy TLS.  The spatial extent of the gap fluctuation sets the frequency of the qTLS, which means that the density of states (DOS) of these objects is very non-uniform (in contrast to conventional TLS in glasses). We predict a maximum in the qTLS DOS around 6-10 GHz, the frequency range of most qubits (transmons).
Relatively shallow traps also imply that even a modest temperature can reshuffle the QPs. 
We expect the underlying qualitative physics of charge defects formed by trapped QPs to be applicable to most materials, including
superconductors in the dirty limit, such as Al used for qubits, but the relevant frequency ranges may differ somewhat due to different material parameters.

This model is supported by a series of experimental observations from detailed mapping of parameter fluctuations in high-Q frequency tunable superconducting NbN resonators \cite{sumedh2019}. 
Such resonators allow us to map individual TLS in energy and trace them in a broad range of external parameters, such as magnetic or electric field \cite{lisenfeld2019}.
Through analysis of a large number of individual TLS we find:
(i) most qTLS are highly coherent, some with detected linewidths down to $\lesssim 70$ kHz, the resonator linewidth. (ii) The qTLS landscape irreversibly reshuffles at very modest temperatures of $\sim 300$ mK,  (iii) The qTLS DOS appears strongly suppressed at lower frequencies. And (iv) tracking a qTLS in magnetic field reveals a qTLS energy that scale in a way similar to the expected scaling of $\Delta$.
In particular, the presence of a low energy scale  $\lesssim1\text{K},$ is inconsistent with conventional TLS theory in which all scales
are set by chemical energies \cite{phillips}. The low energy scale is instead natural for TLS formed by QPs trapped in regions of locally smaller gap. 
We start by outlining the main features of the model that explains these observations.

\addtocontents{toc}{\protect\setcounter{tocdepth}{-1}}
\section*{Results}
To understand the emergence of qTLS from a fluctuating $\Delta = \Delta_0 + \delta\Delta({\bf r})$
we consider a general model that assumes gaussian
fluctuations of the gap, $\left\langle \delta\Delta({\bf r})\delta\Delta({\bf r'})\right\rangle =g\delta({\bf r}-{\bf r'})$, where $g\simeq \delta \Delta\Delta_0\xi_0^2$ is the strength of fluctuations, and $\xi_0$ the BCS coherence length. 
The exact origin of these fluctuations is not relevant
for the conclusions. The density of subgap states is determined by
 rare fluctuations of the gap that reduce the QP 
energy sufficiently below the gap edge in a uniform material. 
The key point is that such fluctuations  can create traps with a shape and depth that allow them to effectively trap QPs.
As the QPs near the gap edge move with momenta close to the Fermi momentum $p_{F}$, it follows that trapped QPs will have similar momenta, which can point in an arbitrary direction.
This freedom results in the optimal trap being very anisotropic, with a shape elongated along the direction
of $p_{F}$. In Fig. \ref{fig:sketch}c we show the typical shape of such a trap obtained through numerical simulations.
In general, the probability to find a fluctuation of depth E scales exponentially with the area $A$ of the fluctuation, $P\sim \exp{(-E^2A)}$. Efficient traps thus have small area, however, a too small dimension along the direction of $p_F$ will forbid the formation of a bound state due to the large kinetic energy of the QP. Therefore optimal traps do not favor isotropic fluctuations. 
The wave function of the QP oscillates quickly along the direction of $p_F$. The
ground and excited states in the trap differ in the number of oscillations:
since the total number of oscillations is large it is not surprising
that each trap typically contains more than one bound state.

\begin{figure*}[t!]
\includegraphics[scale=0.63]{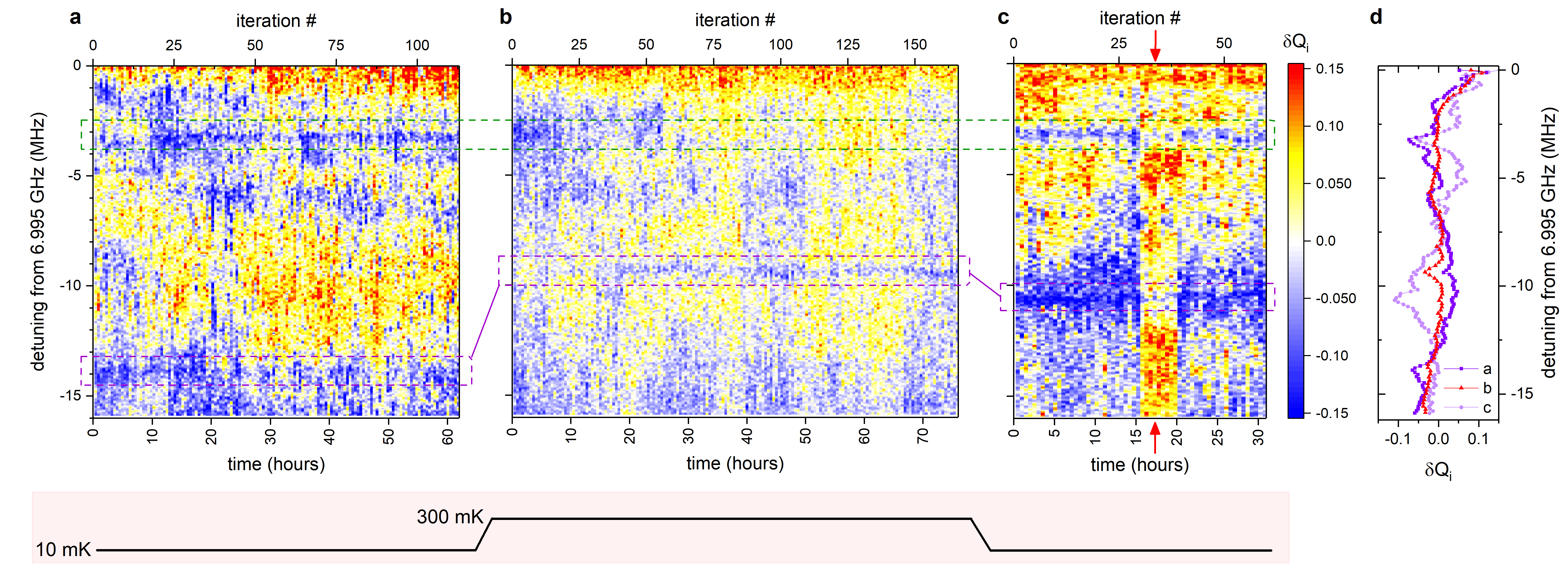} \caption{Spectral \& temporal fluctuations of the internal quality factor and
spectral reconfiguration of TLS. (a) Initial measurement taken at
10 mK and $\langle n\rangle\approx40$. For each pixel the resonator
line-shape is measured and fitted to extract the internal quality
factor $Q_{i}$. 
The measurement is repeated as the resonator is tuned in frequency (the typical frequency resolution is $\approx 50$ kHz), this measurement
vs detuning is then repeated over time. The color scale shows the normalised
variation $\delta Q_{i}=(Q_{i}-\langle Q_{i}\rangle)/\langle Q_{i}\rangle$. 
(b) same measurement performed after
(a) with the temperature
set to 300 mK. (c) the same measurement was repeated again after returning
to 10 mK. The red arrows indicate a broad TLS jumping into the measurement window. Dashed boxes indicates two strong
TLS. For repeated experiments see \cite{suppl}. d) The time-averaged data from panels (a)-(c) respectively.}
\label{fig:spectraltemporal} 
\end{figure*}

Due to the presence of excited bound states, a QP in such a
trap forms an effective TLS with typical energy splitting $\hbar\omega_{\text{qTLS}}=\zeta\left|\widetilde{E}\right|$
where $\widetilde{E}$ is the energy of the trapped QP as measured
from the gap edge and  $\zeta\approx0.7$ is found from theory \cite{suppl}. 
One consequence of these traps is that, in contrast to conventional TLS, the DOS of qTLS is not constant.
A large $\omega_{qTLS}$ implies very deep traps that are exponentially
rare. We find the density
\begin{equation}
\rho_{\text{qTLS}}(\omega)\simeq \nu(\widetilde{E}) = \frac{1}{\xi_{0}^{2}\Delta_{0}}\text{exp}\left[-\frac{\eta}{g}\frac{\xi_{0}^{2}\Delta_{0}^{2}}{\sqrt{k_{F}\xi_{0}}}\left(\frac{\hbar\omega}{\zeta\Delta_{0}}\right)^{5/4}\right],
\end{equation}
where $\eta\approx 10$. Likewise, small $\omega_{qTLS}$ are suppressed since
QPs in shallow traps anihilate each other efficiently, leading to 
\begin{equation}
\rho_{\text{qTLS}}(\omega) \leq \nu_0(\widetilde{E}) = \frac{\sqrt{2k_{F}\xi_{0}}}{\xi_{0}^{2}\Delta_{0}[ \text{ln}(\tau_{\text{exp}} \Delta_0/\hbar) ]^2}\left({\frac{\hbar\omega}{\zeta\Delta_{0}}}\right)^{3/4},
\end{equation}
in the absence of QP generation. Here $\tau_{\text{exp}}$ is an experimental timescale.

Due to the spatial asymmetry, the transition between the lowest and 
first excited state in each trap is expected to have a significant electric dipole
moment, resulting in a strong interaction with quantum devices.
In this picture the TLS is formed in a single well, in contrast to the more conventional double-well picture.

In similar films  $\delta\Delta$ constitute 10\% of the total gap \cite{carbillet2019, lemarie2013, Liao2019}, which gives in our devices $\delta\Delta\approx2.5$ K. 
At $T=10$ mK tunneling out of such a local minima is exponentially suppressed, however, at temperatures approaching fractions of $\delta\Delta$ there is a finite probability that the QP escapes and either finds another minima or recombines, resulting in a typical escape temperature of the order of $ 200-300$ mK; each event resulting in a different TLS landscape.

The unusual shape of the
traps and the presence of multiple bound states in each trap distinguish clean superconductors from the dirty limit  \cite{Larkin_Ovchinnikov, Lamacraft_Simons, Silva_Ioffe}. In the latter a QP is scattered many times by defects
while moving in a particular direction inside the trap, making
the formation of elongated traps impossible. 
The physics of  qTLS traps is  similar to the well studied phenomena of the appearance of states with negative energy
in disordered conductors \cite{Halperin_Lax66,Cardy1978}, with two important differences: electrons at the bottom of the band in a disordered conductor have much smaller momenta, and the wave functions do not have zeros.

\begin{figure*}[t!]
\includegraphics[scale=0.54]{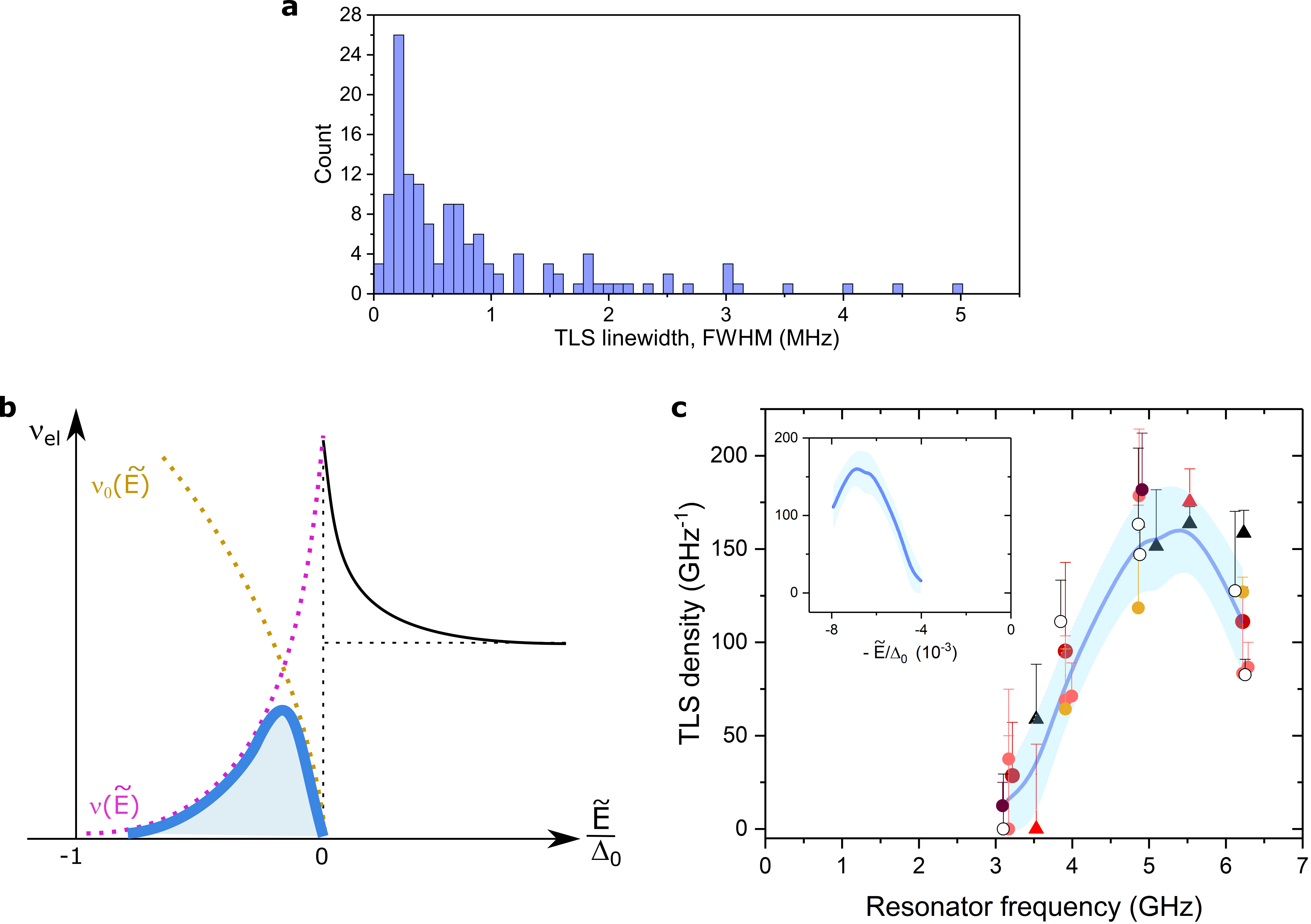} \caption{qTLS Properties. a) Histogram of extracted TLS linewidths (FWHM) obtained
at 10 mK across a series of cooldowns and different resonators in
the frequency range 3-7 GHz.  b)  Expected scaling of the qTLS density of states (blue solid region) bounded by  $\nu(\widetilde{E})$ and $\nu_0(\widetilde{E})$ (see text for details). Not to scale. c) Number of strongly coupled TLS as a function of frequency. The data
points are obtained by counting the number of TLS present in a spectral
and temporal map. Data taken on two different samples (circles and triangles) in seven separate cooldowns (different colors). Duplicate markers at the same frequency are before/after thermal reconfiguration. Multiple markers from the same resonator have been offset somewhat in frequency for better visibility. 
Error bars indicate miscounting the number of TLS by one, e.g. if one
TLS happens to be present just outside the measurement range. The solid line is the average of observations, bounded by the standard deviation. The inset shows the same data plotted against the corresponding trap depth from $\Delta_0$. All data obtained at 10 mK and $\langle N\rangle<100$.}
\label{fig:linewidths} 
\end{figure*}

We now turn to the experimental work where our observations can be explained by this model.
To be able to study individual TLS and spectral and temporal fluctuations
of superconducting resonator parameters we use kinetic inductance
tunable superconducting resonators with centre frequencies $f_0$ in the
range 3-7 GHz, described in detail in Ref. \cite{sumedh2019}. The resonators
are cooled down to a base temperature of 10 mK in a well filtered
dilution refrigerator. A small dc current ($<1$ mA) is applied to change the kinetic inductance and tune $f_0$.
We measure the transmitted microwave signal, $S_{21}$, using a vector network analyser and 
extract the resonator frequency and quality factors Q. In some of the
experiments we also utilise a gate electrode mounted in the lid of
the sample enclosure, such that we can affect the TLS energy through
the applied gate voltage $V_g$: $E_{{\rm TLS}}=\sqrt{\epsilon_{0}^{2}+\gamma^2 V_{g}^{2}}$ \cite{bilmes2019,lisenfeld2019}. Here $\epsilon_{0}$ is the TLS minimum energy and $\gamma$ the TLS voltage coupling strength.

In Fig. \ref{fig:spectraltemporal}a we show the relative variation
of the internal quality factor $\delta Q_{i}=(Q_{i}-\langle Q_{i}\rangle)/\langle Q_{i}\rangle$, 
as we tune the resonator frequency. The measurement is continously
repeated over several days to produce a spectral and temporal map
of the variations in loss. This reveals large drops in $\delta Q_{i}$: at detunings of 
$3.5$ and $14$ MHz (from 6.995 GHz) TLS dips remain in the same place whereas at
other frequencies both switching-like behaviour and drift can be observed.
We then raise the temperature of the cryostat to 300 mK and repeat
the same measurement (Fig. \ref{fig:spectraltemporal}b), and return to 10 mK (Fig. \ref{fig:spectraltemporal}c). We observe that a strong TLS 
remains stable over more than 60 hours in each individual measurement,
but thermal cycling to a mere 300 mK makes another TLS  appear at a different frequency.
This is quite remarkable as the temperature is increased to a value
smaller than the energy level splitting of the TLS itself (7 GHz $\approx$
350 mK), and certainly not expected from conventional TLS-glass physics
in which all scales are set by chemical energies \cite{phillips}.
Separate high power measurements show that these dips are saturable, and not caused by e.g. local variations
in background transmission. We have repeated the same experiment but visiting multiple temperatures
on the way to 300 mK (see  \cite{suppl} for additional data) which confirms that reconfiguration only
happens near 300 mK.

\begin{figure*}[t!]
\includegraphics[scale=0.68]{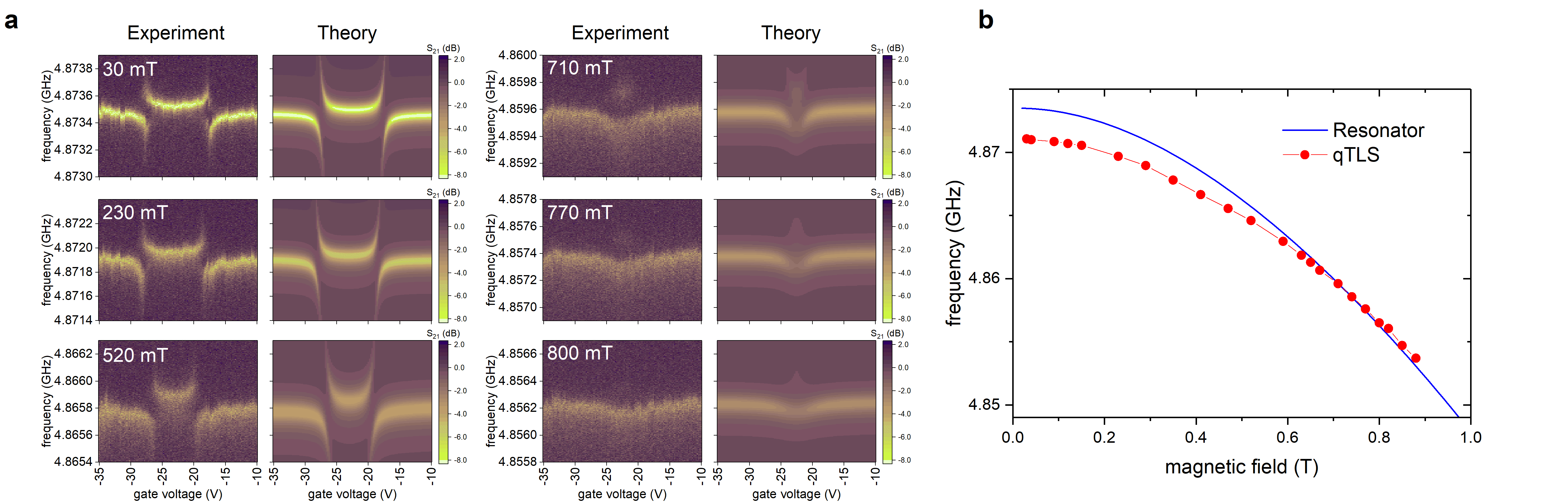} \caption{A single strongly coupled qTLS in magnetic field. a) Transmission
spectrum of the resonator strongly coupled to a qTLS at selected magnetic
fields, as a function of applied electrostatic gate voltage, showing
the coherent coupling of a single qTLS to the resonator. b) The resonance
frequency as a function of magnetic field (solid blue line) together
with the fitted qTLS minimum energy $\epsilon_{0}$ (red markers).
Other extracted parameters are qTLS-resonator coupling strength $g_0=270$
kHz, damping rate $\gamma_{{\rm qTLS}}=200$ KHz and $\gamma=10$ MHz/V, consistent with $p_0\approx 1$ $\rm e\AA$ in our geometry.}
\label{fig:tlsbfield} 
\end{figure*}

Occasionally we observe very dissipative TLS with linewidths in excess of several MHz in our measurement window (red arrow in Fig. \ref{fig:spectraltemporal}c). Here we concentrate instead on the much more coherent TLS that we consistently observe, the narrowest with
line-widths down to $\sim70$ kHz \cite{suppl}, limited by the readout resonator line-width. In Fig. \ref{fig:linewidths}a we show the histogram of
linewidths of the observed TLS, which peaks at values corresponding to coherence times in excess
of several microseconds, much longer than what has been reported for individually resolved TLS in qubits \cite{lisenfeld2019,lisenfeld2015}. The energy leakage caused by  such highly coherent TLS could pose a significant challenge for quantum error correction \cite{Brown2019-1}.
Naturally, qTLS interact with conventional TLS and slow fluctuators that result in parameter fluctuations.

We now turn to the qTLS DOS. Fig. \ref{fig:linewidths}b shows the expected frequency dependence of the qTLS DOS, given by the bounds of eq. (1) and (2).
This is in good agreement with the number of strongly coupled qTLS
that we observe in resonators with different $f_0$, shown in Fig. \ref{fig:linewidths}c; data obtained across several cooldowns and devices. A similar dependence was obtained by tuning the TLS energy by an applied electrostatic field, see \cite{suppl}. 
In contrast, for conventional non-interacting TLS in amorphous materials we expect
a DOS $P(E)={\rm {const}}$ \cite{phillips}, while for interacting
TLS we expect a weak pseudogap given by $P(E)=E^{\mu}$ \cite{faoro2015, moshe2020}.  $\mu\approx0.2$ is typically found experimentally \cite{burnett2014,degraaf2018}, which would for $f_{0}=3-6$ GHz only result in a $\sim10\%$
variation in the TLS DOS. 
As expected, we find through the temperature dependent shift of $f_0$ which samples all TLS, including the more numerous conventional glassy TLS,  that the intrinsic loss tangent fall within $F\tan\delta_{i}=0.7\pm0.1 \cdot10^{-5}$,
with no observed dependence on $f_0$ \cite{suppl}.

The data in Fig. \ref{fig:linewidths}b has not been
scaled to the surface area occupied by each resonator, which would yield an even stronger non-linear
dependence ($A\propto f_{0}$). If we make the assumption
that the observed qTLS uniformely occupy the surface of the superconductor we get for $f_0=4.9$ GHz an observed density of qTLS of $0.004$ GHz$^{-1}$ ${\rm \mu}$m$^{-2}$, or $1$ GHz$^{-1}$ per $(\sim3000\xi_0)^2$.
In our devices the distance between electrodes is small, resulting
in significant electric fields away from the immediate vicinity
of metal edges. In fact, the maximum observed TLS coupling strengths of
up to $100-200$ kHz are obtained for electric dipoles of size $p_0\sim 1$
${\rm e\lyxmathsym{\AA}}$ (the expected scale of a qTLS dipole moment \cite{suppl} and commonly found also for conventional TLS \cite{lisenfeld2019, sarabi2016}) across a significant part of the device
surface. Near metal edges we would expect coupling strengths well exceeding
1 MHz \cite{suppl}. This we do not observe, implying instead that the qTLS are detected across
most of the device surface.
The above density is much smaller than the total density of TLS we extract
from $F\tan\delta_{i}$. We find a surface density
of $\rho_{{\rm \Delta,A}}\approx50$ GHz$^{-1}$ ${\rm \mu}$m$^{-2}$
of weakly coupled TLS which are mainly responsible for the background
loss and noise \cite{suppl} previously studied in detail \cite{burnett2014, degraaf2018}. These conventional TLS coexist with the qTLS and constitute the much broader background of loss and fluctuations. However,
since the qTLS are located at the metal surface they interact
weakly with the conventional glassy TLS bath which is mainly situated
some distance away, e.g. in the oxide on the superconductor and adsorbants
on top, resulting in higher qTLS coherence.

Finally, to determine if there indeed is a link between the qTLS and
superconductivity we measure a strongly coupled qTLS 
in an applied
magnetic field $B$ parallel to the superconducting film, shown in Fig.
\ref{fig:tlsbfield}a. As expected, we observe a quadratic suppression
of $f_0$ owing to the kinetic inductance 
increasing as $\Delta$ is suppressed. At the same time
we track the anti-crossing with the qTLS and extract the minimum qTLS
energy $\epsilon_0(B)$. As shown in Fig. \ref{fig:tlsbfield}b
$\epsilon_0$ is also suppressed in magnetic field on a
scale comparable to the weak (few \%) suppression of $\Delta$. 
We also note that the conventional TLS bath remains unaffected up to moderate fields \cite{degraafAPL2018}.

High-resolution spectral mapping of TLS in qubits has only been performed in the range $ 5-6$ GHz \cite{klimov2018}.  Below $3$ GHz qubits
in other studies were limited by other loss mechanisms or statistics are insufficient \cite{nguyen2019,quintana2017,yan2015}.
Only Ref. \cite{quintanathesis} hints at a lower TLS density in Al qubits below $\simeq 3$ GHz,
suggesting that the mechanism
responsible for the formation of qTLS could indeed be quite common, and motivating the need to explore lower frequencies.
The critical parameter for the DOS, the strength of fluctuations $g$, is to the best of our knowledge not known for thin Al films. However, it may even be enhanced due to thickness fluctuations contributing to $\delta \Delta$ \cite{ivry2014,aldelta} and the upper frequency for qTLS is bounded by a smaller $\Delta_0$.  
We  note that in the dirty limit the details of the theory developed here is not strictly valid. However, the general mechanism would still apply, but the optimal shape of fluctuations is likely to be different.

We now turn to the implications that these qTLS states have for a range of superconducting technologies and for the coherence of qubits. A typical sample of $\sim 1$ cm$^{2}$ size is hit by a high energy
cosmic ray once in every $\sim10$ s, creating a shower of phonons that quickly relax to low frequencies where their relaxation time becomes long. These
phonons produce QPs well above $\Delta$. 
Eventually these QPs fall into the traps discussed here where they remain
essentially forever at low temperatures. Such localization of QPs prevents their recombination and equilibration. Those close to a qubit can result in its excitation or relaxation, and hence the QP localization results in a
long period of qubit performance degradation. As this degradation affects all qubits on a
sample it becomes a catastrophic event for quantum computation, a conclusion supported by recent data
on charge noise in qubits \cite{mcdermott}
 and by the studies of QP bursts in granular Al
resonators \cite{Henrique2019} and kinetic inductance detectors \cite{karatsu2019}. 

Identifying and eliminating strongly coupled TLS is crucially important for scaling-up of superconducting quantum computing.  
All the mechanisms that lead to gap fluctuations are difficult to control in thin films used for resonators and qubits, but our results suggest these must be carefully addressed in order to achieve this goal.  

%

\section*{Acknowledgements}
We thank R. McDermott and M. Gershenson for fruitful discussions. Samples were fabricated in the nanofabrication facilities of the department of microtechnology and nanoscience at Chalmers university. This work was supported by the UK Department of Business, Energy and Industrial Strategy (BEIS), the EU Horizon 2020 research and innovation programme (grant agreement 766714/HiTIMe), the Swedish Research Council (VR) (grant agreements 2016-04828 and 2019-05480), EU H2020 European Microkelvin Platform
(grant agreement 824109), and Chalmers Area of Advance NANO/2018.
J.J.B acknowledges financial support from the Industrial Strategy Challenge Fund Metrology Fellowship as part of BEIS.

\let\oldaddcontentsline\addcontentsline
\renewcommand{\addcontentsline}[3]{}

\let\addcontentsline\oldaddcontentsline
\clearpage
\onecolumngrid
\appendix
\section*{\Large{Supplemental material}}\thispagestyle{empty}\vspace{20mm}
\renewcommand\thefigure{\thesection S\arabic{figure}}    
\setcounter{figure}{0} 
\setcounter{equation}{0} 
\renewcommand\theequation{S\arabic{equation}}
\maketitle
\tableofcontents
\addtocontents{toc}{\protect\setcounter{tocdepth}{3}}

\clearpage
\section{Derivation of the qTLS density of states\label{subsec:Derivation-of-the}}
Even in very clean superconducting films the gap varies significantly
on the scale of superconducting coherence length, $\xi_{0}$. For
instance, in ultra clean NbN thin films spatial fluctuations of the
order of $\sim10\%$ in $\Delta$ have been reported in \cite{carbillet2019}.
These fluctuations can be caused by the inhomogeneous pair breaking
by weak magnetic impurities on the surface or by inhomogeneity of
the density of states in these films, characterized by very low carrier
density. These fluctuations provide traps for quasiparticles with
the energies below the average gap, $\Delta_{0}.$ Below we compute
the density of quasiparticle states in these traps and their typical
spectrum in each trap. We find the a typical trap contains, apart
from the ground state, a few localized excited states. We shall assume
that gap fluctuations are much smaller than the gap itself, $\delta\Delta\ll\Delta,$
so that the states are formed at energy $\Delta-E$ with $E\ll\Delta$
which is the experimentally relevant case. 

Without the loss of generality we can assume that spatial fluctuations
in $\Delta$ are described by a white Gaussian noise potential that
is characterized by Gaussian statistics with auto-correlation function:

\begin{equation}
\begin{array}{c}
W({\bf r}-{\bf r'})=\langle\delta\Delta({\bf r})\delta\Delta({\bf r'})\rangle=g\delta({\bf r}-{\bf r'})\\
P(\delta\Delta)=\frac{1}{\sqrt{2\pi g}}e^{-\frac{1}{2g}\int\delta\Delta^{2}dxdy},
\end{array}\label{eq:SP}
\end{equation}
where $\delta(r)$ is the dirac delta function.  As we show below, the relevant scale for the quasipartcle traps is much larger than $\xi_0$, justifying the  $\delta$-correlation.
We now derive the density of subgap states, $\nu(E)$ and the shape
of the optimal fluctuation of the gap that produces them. We focus
on the low energy tail. Throughout we use $\hbar=1$. The starting point is the Bogoliubov de Gennes
(BdG) equations:

\[
\begin{array}{c}
H_{0}u_{n}({\bf r})+\Delta({\bf r})v_{n}({\bf r})=E_{n}u_{n}({\bf r})\\
-H_{0}v_{n}({\bf r})+\Delta^{*}({\bf r})u_{n}({\bf r})=E_{n}v_{n}({\bf r}),
\end{array}
\]
where ${\bf r}={\bf \hat{x}}x+{\bf \hat{y}}y$ (we consider the 2D problem,
generalization to 3D is straightforward). $H_{0}=\frac{1}{2m}(-i\nabla_{{\bf r}})^{2}-E_{F}$,
$E_{F}=\frac{k_{F}^{2}}{2m}$ is the Fermi energy, $\Delta({\bf r})=\Delta_{0}+\delta\Delta({\bf r})$
and $\xi_{0}=\frac{v_{F}}{\Delta_{0}}$ denotes the coherence length
of the superconductor. If we regard $u_{n}({\bf r})$ and $v_{n}({\bf r})$
as the upper and lower components of a ``spinor'' $\psi_{n}({\bf r})$,
we can write the effective Hamiltonian acting on this spinor as a
matrix: $H=H_{0}\sigma_{z}+\Delta\sigma_{x}$. In the limit $H_{0}\ll\Delta$, we can use perturbation theory to reduce the Hamiltonian $H$ to 

\[
\widetilde{H}=\Delta+H_{0}\frac{1}{\Delta}H_{0}.
\]
The BdG equations then reduce to:

\begin{equation}
\left[\frac{H_{0}^{2}}{\Delta_{0}}+\delta\Delta({\bf r})\right]v_{n}=\widetilde{E}_{n}v_{n}\quad\text{where\ensuremath{\quad\widetilde{E}_{n}=E_{n}-\Delta_{0}<0} }.
\end{equation}
The BdG equations are valid on the atomic scale and therefore the spinor
wave functions $\nu_{n}(r)$, which vary on the length scale set by
$k_{F}^{-1},$ contains more information than needed. It is convenient
to eliminate these irrelevant degrees of freedom by replacing the
BdG equations by their quasiclassical limit. For this purpose one
writes the spinor wave function $\nu_{n}({\bf r})$ as a rapidly oscillating
phase factor (which changes on the atomic length scale) times a slowly
varying amplitude (which changes on a length scale set by the coherence
length), i.e. $v_{n}({\bf r})\approx\varphi_{n}(x,y)e^{ik_{F}{\bf \hat{x}\cdot{\bf r}}}$.
In this quasiclassical approximation, the quasiparticles are moving
along trajectories that are straight lines. We then rewrite the BdG
equations neglecting higher derivatives in the $x$ direction 
\[
\left|\frac{\nabla_{x}^{2}\varphi_{n}(x,y)}{k_{F}{\bf \hat{x}}\nabla_{x}\varphi_{n}(x,y)}\right|\sim\left(k_{F}\xi_{0}\right)^{-1}\ll1.
\]
We find that:
\begin{equation}
H_{0}\approx\left[-iv_{F}\frac{\partial}{\partial x}-\frac{1}{2m}\frac{\partial^{2}}{\partial y^{2}}\right]\label{eq:Kin},
\end{equation}
and the BdG equations reduce to:

\begin{equation}
\left\{ \frac{1}{\Delta_{0}}\left[-iv_{F}\frac{\partial}{\partial x}-\frac{1}{2m}\frac{\partial^{2}}{\partial y^{2}}\right]^{2}+\delta\Delta({\bf r})\right\} \varphi_{n}(x,y)=\widetilde{E}_{n}\varphi_{n}(x,y)\label{eq:Bdg}.
\end{equation}

After these simplifications, the problem becomes formally similar
to the problem of finding the tail in the density of states in disordered
conductors \cite{Halperin_Lax66,Cardy1978} or the density of subgap
states due to gap variation in dirty superconductor \cite{Larkin_Ovchinnikov}. The important difference with these works is, however, the very
anisotropic form of the kinetic energy in \cite{Larkin_Ovchinnikov}. 

For the optimal fluctuation all the terms in Eq. \ref{eq:Bdg} are of the same order. This implies that the optimal fluctuation has the size:
\begin{equation}
\begin{array}{c}
x_{0}=\sqrt{\frac{\Delta_{0}}{\widetilde{E}_{n}}}\xi_{0}\\
y_{0}=\frac{1}{\sqrt{2}\sqrt{k_{F}\xi_{0}}}\sqrt[4]{\frac{\Delta_{0}}{\widetilde{E}_{n}}}\xi_{0}
\end{array}\label{eq:Scales-1}
\end{equation}
in $x$ and $y$ directions respectively. Notice that the scales of
the two coordinates are different, i.e. $x_{0}\gg y_{0}$, resulting
in an optimal gap fluctuation with no spherical symmetry (see Fig. 1c in the main manuscript).

By using Eq. \ref{eq:SP}, we find that the probability to find the
fluctuation $\delta\Delta\sim\tilde{E}$ with spatial extend $x_{0}$
and $y_{0}$ scale as: $\text{ln}P(\tilde{E})=-\frac{\eta'}{2\sqrt{2}g}\frac{\xi_{0}^{2}\Delta_{0}^{2}}{\sqrt{k_{F}\xi_{0}}}\left(\frac{\widetilde{E}}{\Delta_{0}}\right)^{5/4}$.
For a 3D problem, similar reasoning gives: $\ln P(\widetilde{E})=-\frac{\eta'}{2\sqrt{2}g}\frac{\xi_{0}^{2}\Delta_{0}^{2}}{k_{F}\xi_0}\frac{\widetilde{E}}{\Delta_{0}}$. 

In order to find the exact value of the coefficient $\eta'=2\sqrt{2}\eta\sim O(1)$
we need to determine the optimal gap fluctuation that dominates the
density of states at energy $\tilde{E}.$ We follow the derivation
given in Ref. \cite{Halperin_Lax66} of the mian text. By introducing the adimensional
coordinates:

\[
\begin{array}{c}
\bar{x}=x_{0}^{-1}x\\
\bar{y}=y_{0}^{-1}y,
\end{array}
\]
and the scaled wavefunction $\widetilde{\varphi}(\bar{x},\bar{y})=\sqrt{\frac{g}{\widetilde{E}}}\varphi(\bar{x},\bar{y})$,
we find that Eq. \ref{eq:Bdg} reads:

\begin{equation}
\left[-i\frac{\partial}{\partial\bar{x}}-\frac{\partial^{2}}{\partial\bar{y}^{2}}\right]^{2}\widetilde{\varphi}(\bar{x},\bar{y})-\widetilde{\varphi}^{3}(\bar{x},\bar{y})-\widetilde{\varphi}(\bar{x},\bar{y})=0\label{eq:solnum-1},
\end{equation}
and the coefficient:
\begin{equation}
\eta'=\int\left|\widetilde{\varphi}(\bar{x},\bar{y})\right|^{4}d\bar{x}d\bar{y}\label{eq:eta}.
\end{equation}

The spectrum operator in Eq. \ref{eq:solnum-1} determines the
presence (or absence) of excited bound states in the optimal gap fluctuation.
Eq. \ref{eq:solnum-1} is a fourth order non linear differential equation
that we can solve numerically. The solution $\widetilde{\varphi}(\bar{x},\bar{y})$
is illustrated in Fig. 1c in the main manuscript and it represents the
optimal trap potential. To solve the non-linear differential equation (\ref{eq:solnum-1}) we proceed in the following way. The minimization of a generic quadratic functional $F(\phi)$ with the constraint $\Lambda(\phi)$ defined by:
\begin{equation}
F=\int \phi^* M \phi dx dy \qquad \Lambda=\int |\phi|^4 dx dy 
\end{equation} 
is given by the solution of the Lagrange equation 
\begin{equation}
M\phi + \lambda |\phi|^2 \phi = 0
\end{equation}
where $\lambda$ is the Lagrange multiplier. The latter equation can be reduced to equation (\ref{eq:solnum-1}) by the rescaling of the function $\phi$. Thus, minimizing the functional $F(\phi)$ with appropriate choice of the operator $M$ and rescaling the result we get the solution of equation (\ref{eq:solnum-1}). The minimization procedure is numerically stable in contrast to the direct solution of the non-linear equation (\ref{eq:solnum-1}).

The non spherical symmetry is clearly seen
and has important consequences for the estimates of the bound states
in the typical well.
We find that in the optimal trap, beside the ground state $\widetilde{E}$,
there is a distintive excited bound state at energy $\widetilde{E}(1-\zeta)$
with $\zeta\approx 0.7$. As a consequence when a quasiparticle is trapped
in the optimal gap fluctuation, a qTLS with frequency $\omega_{qTLS}\sim\zeta\left|\widetilde{E}\right|$
can be formed. 

As a result of the numerical calculation, we find that the adimensional
parameter given in Eq. \ref{eq:eta} is $\eta'\approx 28$.

Two bound states in the well form the qTLS. 
We now estimate the density of qTLS, $\rho_{qTLS}(\tilde{E})$.
For energies into the low-energy tail, the density of states of the qTLS is given by the electron density of states:
\begin{equation}
\nu(\widetilde{E})\approx\frac{1}{\xi_{0}^{2}\Delta_{0}}\text{exp}\left[-\frac{\eta'}{2\sqrt{2}g}\frac{\xi_{0}^{2}\Delta_{0}^{2}}{\sqrt{k_{F}\xi_{0}}}\left(\frac{|\widetilde{E}|}{\Delta_{0}}\right)^{5/4}\right]\label{eq:densitfin}.
\end{equation}
The reason is that at such a low energies, the qTLS are well localized
into the optimal gap fluctuations. However, traps with value of energies
$\widetilde{E}\to0$ are very shallow and a trapped quasiparticle
can tunnel out of the well and, depending on the time of experiment
$\tau_{\text{exp}}$, eventually find a partner and annhilate. 
If this happens, the qTLS formed in such a shallow trap disappears.
We therefore expect that the density of qTLS varies strongly
depending on the qTLS frequency. In particular, one would expect that at
very high frequencies qTLS are exponentially rare, the majority of
qTLS are centered around some typical frequency, and at eneriges corresponding
to very shallow traps there are again no qTLS. This is in agreement with
what has been observed in the experiment (see Fig. \ref{fig:linewidths}c in the main manuscript). We can estimate how the density of qTLS decreases at very small
values of $\widetilde{E}$. To this aim, we estimate the tunneling
amplitude of the quasiparticle in the trap:

\begin{equation}
t_{\text{amp}}=|\widetilde{E}|\varphi(\bar{x},\bar{y}),
\end{equation}
and expect that two quasiparticles cannot annhilate if 
\begin{equation}
\tau_{\text{exp}}t_{\text{amp}}\ll1\label{eq:annhi}.
\end{equation}

Since $\varphi(\bar{x},\bar{y})\sim\text{exp}\left[-\left(\frac{x}{x_{0}}+\frac{y}{y_{0}}\right)\right]$,
the condition given in Eq. \ref{eq:annhi} is satisfied if

\begin{equation}
\frac{x}{x_{0}}+\frac{y}{y_{0}}\gg\text{ln\ensuremath{\left(\tau_{\text{exp}}|\widetilde{E}|\right)}}.
\end{equation}

This translate into the fact that each shallow trap is associated
a surface $S=x_{0}y_{0}\text{\ensuremath{\left[\text{ln}\left(\Delta_0\tau_{\text{exp}}\right)\right]}}^{2}$ and
only for trap densities $\nu_{0}(\widetilde{E})<S^{-1}$, the shallow
trap behaves effectively as a qTLS. It is a strightforward calculation
to show that:

\begin{equation}
\nu_{0}(\widetilde{E})=\frac{\sqrt{2k_{F}\xi_{0}}}{\xi_{0}^{2}\Delta_{0}[ \text{ln}(\tau_{\text{exp}} \Delta_0) ]^2}\left({\frac{|\widetilde{E}|}{\Delta_{0}}}\right)^{3/4}.
\end{equation}
As a result one expects that the densities of qTLS at high frequencies
decrease as $\rho_{qTLS}<\nu_{0}(\widetilde{E})$. Fig. 3b in the main manuscript
shows the theoretical expected density of qTLS.

We also note that the subgap states discussed here are 
different from Yu-Shiba-Rusinov states due to strong
magnetic impurities inside the superconductor. Here each impurity
forms a bound state for a QP \cite{Fominov2011}, which is related to Kondo physics
\cite{saveskul2019,proslier2008,heinrich2017,huang2019}.
Impurities with Kondo temperature $T_{K}^{*}\sim\Delta$ produce subgap
states states in the middle of the gap. Morever, for strong magnetic
impurities, energy absorption is due to the tunneling between two subgap
states in two different traps. The only energy scale in this problem
is $\Delta$, in contrast to our experimental observations. 

Finally we make a note on the qTLS electric dipole moment.
As the size of the qTLS trap is on the scale $\sim\xi_0^3$ one would expect a significant dipole moment, however, screening inside the superconductor reduces the effective dipole moment significantly. Any bulk qTLS would have vanishing coupling; only qTLS located at the superconductor surface are expected to couple to external fields. 
However, even in the most favourable situation when a qTLS trap is located at and oriented parallel to the surface of the superconductor the relevant screening length for perpendicular electric fields is on the atomic scale,  $\sim1$ $\rm \AA$. This sets the scale of the dipole moment for both microwave and electric field induced by the electrostatic gate. 
The dipole moment we observe both through the coupling to the microwave field and to the DC gate electrode is consistent with this length-scale.

\clearpage

\section{Device design and fabrication}
The resonator design and fabrication is outlined in detail in \cite{sumedh2019}, and the resonators used here are of exactly the same design but of varying length (frequency). The resonator structure is patterned from a 140 nm Nbn film deposited on a heated 2'' Sapphire substrate using electron beam lithography. The film is etched in a $Ar:Cl_{2}$ reactive ion plasma which ensures sharp sidewalls and prevents lateral under-resist etching. Wire bonds for rf and dc connections are facilitated by $Au$ bonding pads which are deposited in the next step. The resonator (excluding the ground plane) is thinned down to 50 nm in the same $Ar:Cl_{2}$ reactive ion plasma in the final fabrication step. This yields a sheet kinetic inductance of $\sim 4$ pH/$\square$ and places the resonance frequency in the 3-8 GHz band. The sheet kinetic inductance of the ground plane is lower than the resonator film ensuring that the typical frequencies of the ground plane resonances are placed above 8 GHz. Following the final etching step which produces the final film surface residual resist mask is removed in 1165 remover followed by an isopropanol and water rinse. The final step is a soft $O_2$ plasma clean. A second set of samples used to compare the TLS density ustilises a thinner film, 17 nm, with no apparent effect on the qTLS density.

\section{Properties of the superconducting film}
\begin{figure}[h!]
\includegraphics[scale=0.3]{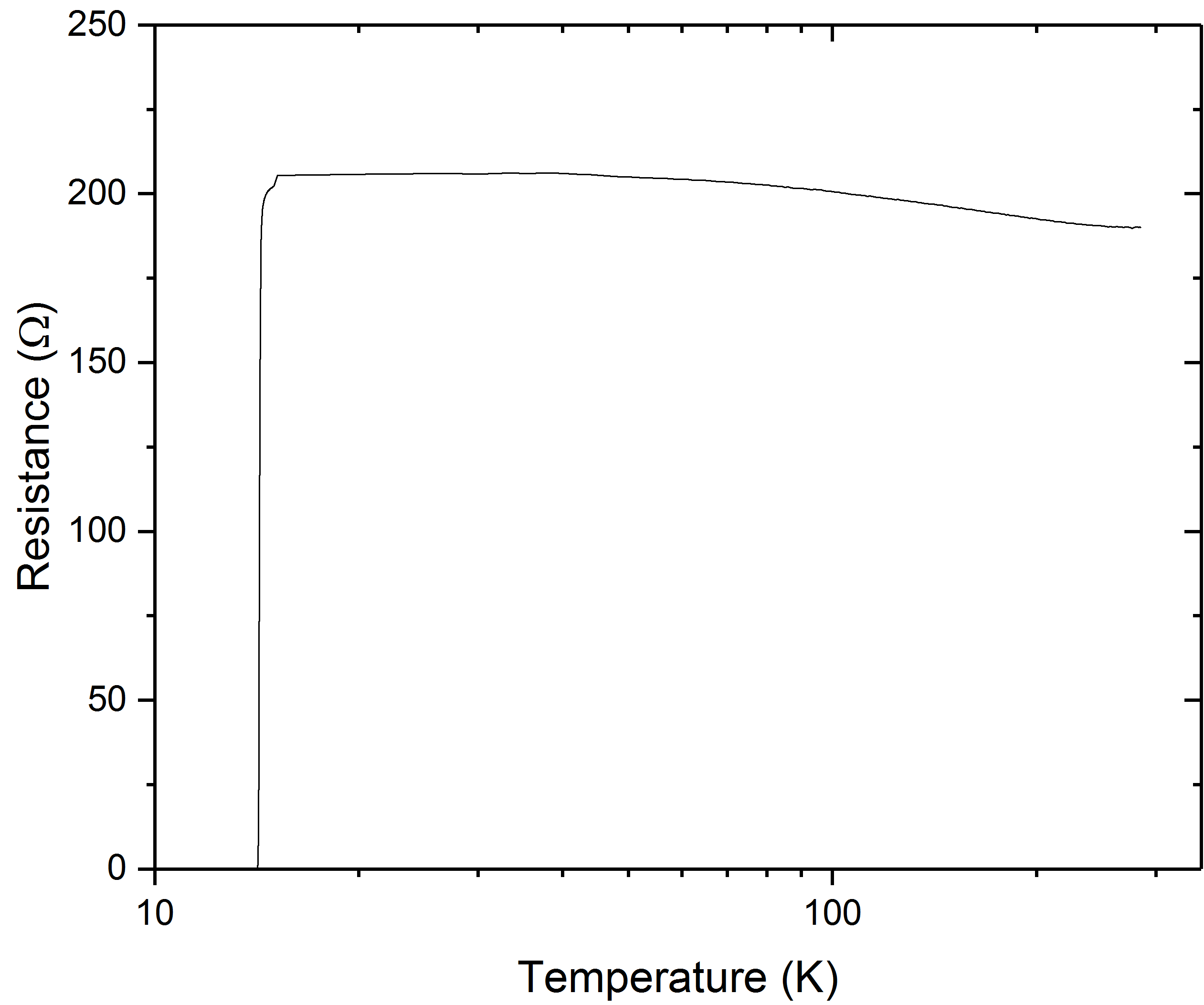} \caption{Resistance versus temperature measured on a 50 nm thick and 6 squares long wire of the NbN film.}
\label{fig:resistivity} 
\end{figure}

Figure \ref{fig:resistivity}
shows the resistivity of the film versus temperature for a typical NbN film deposited in our sputtering system. We find a sheet
resistance of $R_{\square}\approx 35\Omega$ for a 50 nm thick film and
the critical temperature is $T_{c}=14.2$ K and the transition is
sharp with $\Delta T_{c}\approx0.1$ K. The low sheet resistivity and the weak temperature dependence of the resistivity above $T_c$ indicates a relatively clean film. 

\clearpage
\section{Measurement setup}
\begin{figure}[h!]
\includegraphics[scale=0.6]{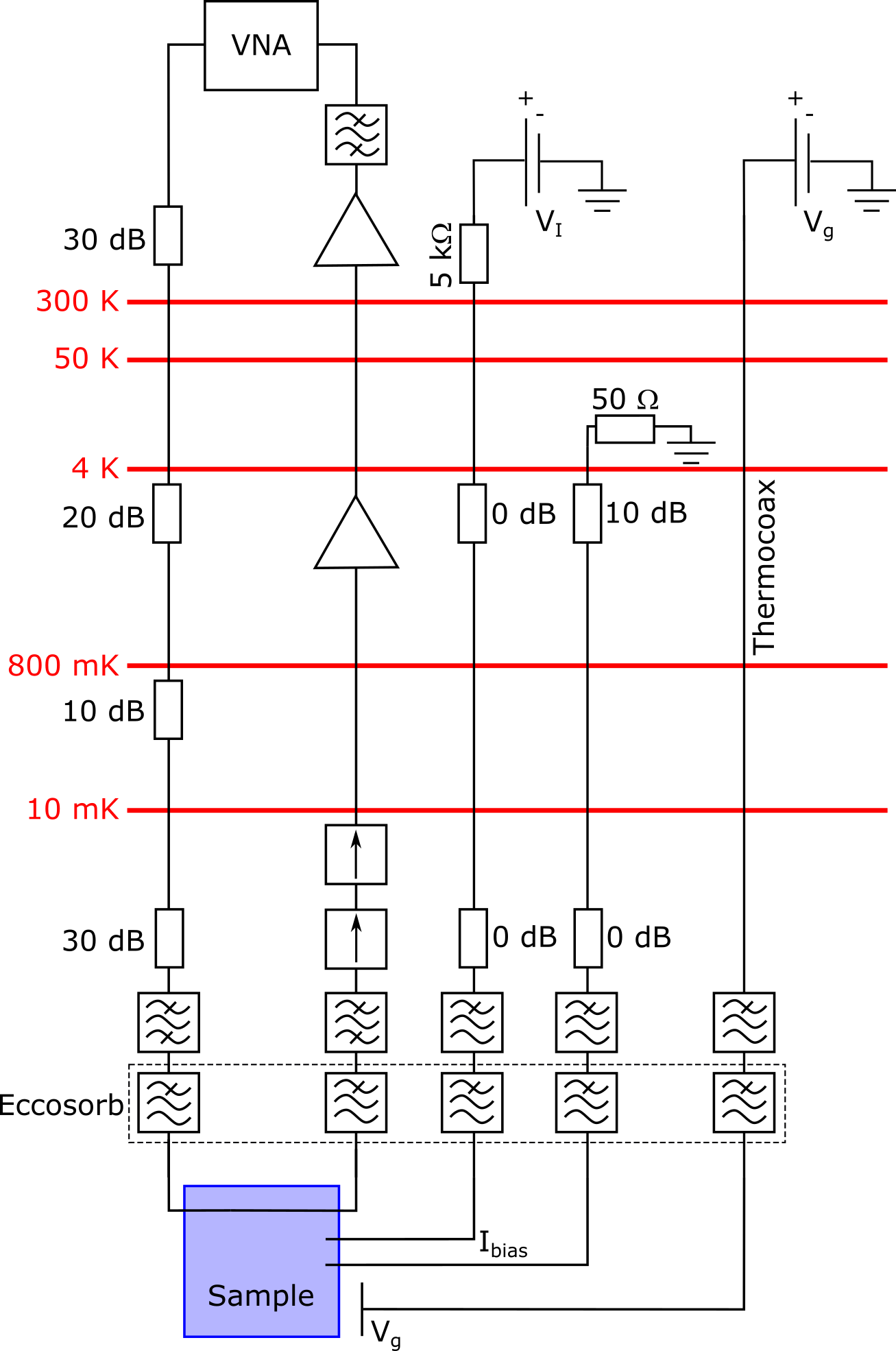} \caption{Schematic showing the main components of the experimental setup.}
\label{fig:setup} 
\end{figure}

Our typical measurement setup is depicted in Fig. \ref{fig:setup}.
The current that is used to tune the resonator frequency is generated
by a voltage source (waveform generator) and converted to
a bias current through a 5 k$\Omega$ resistor. The current is fed
to the sample through un-attenuated low-pass filtered coaxial lines
and back to the 4 K stage of the fridge where it is dissipated in
an attenuator terminated with a 50 $\Omega$ load. At 10 mK all the
measurement and control lines are typically filtered through home-made
eccosorb filters and commercial band- \& low- pass filters as appropriate.
We note that in some experiments (in particular those in the bore of
the magnet) for technical reasons we did not use as elaborate filtering and sample shielding,
however, this appears to have no significant impact on the device
perofrmance or the TLS observed. During magnetic field measurements
the cryogenic isolators on the microwave readout line were also thermalised
on the 800 mK stage to avoid significant stray fields from the magnet,
at the expense of poorer readout constrast due to an increased amount
of thermal photons. 

\clearpage
\section{Loss tangent}
\begin{figure}[h!]
\includegraphics[scale=0.75]{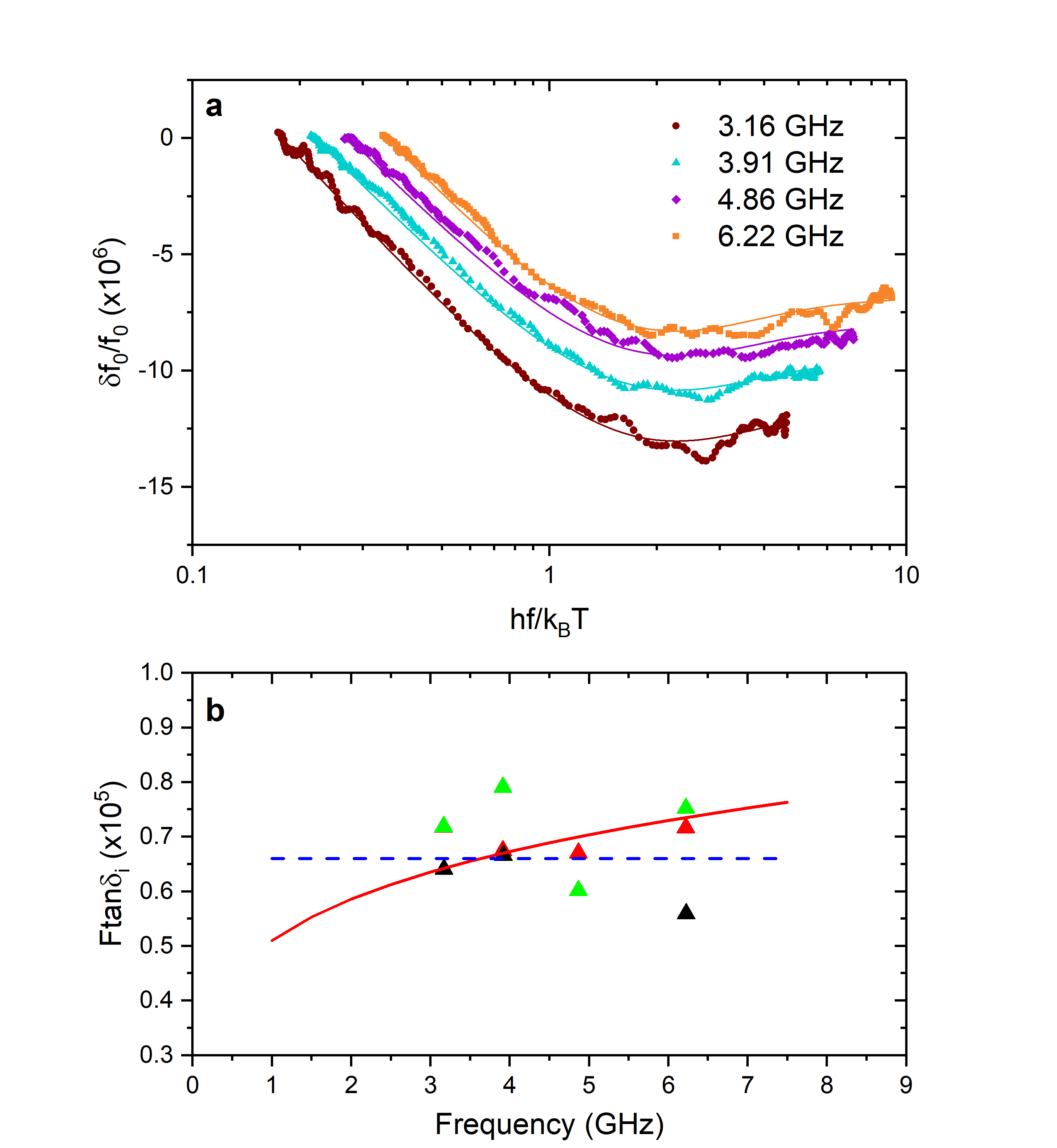} \caption{a) Temperature dependent frequency shift of four resonators measured
simultaneously upon cooling down the sample. Solid lines are fits
to eq. (\ref{eq:tandi}a) for one measurement cooldown. b) The fitted
loss tangents across three cooldowns (different colours) toghether
with the expected scaling due to interacting TLS (solid red line, $\mu=0.2$) and the standard tunneling model result (dashed blue line,  $\mu=0$), see text for details. Error bars from fits are smaller than the symbols.}
\label{fig:tand} 
\end{figure}

We next verify the background loss in our resonators, originating from a large ensemble of weakly coupled TLS (the expected usual dielectric TLS material defects in addition to qTLS), by measuring the intrinsic dielectric loss tangent. The intrinsic loss tangent $F\tan\delta_{i}$ (at zero frequency detuning) of each resonator was determined on multiple cooldowns by sweeping the temperature of the cryostat and measuring
the resonance frequency versus temperature using a VNA. The intrinsic
loss tangent is given by 
\begin{eqnarray}
\frac{\delta f_0(T)}{f_0} & = & F\tan\delta_{i}\bigg[{\rm Re}\bigg(\Psi(\frac{1}{2}+\frac{ f_0h}{2\pi i k_{B}T})
 +\Psi(\frac{1}{2}+\frac{ f_0 h}{2\pi ik_{B}T_{0}})-\ln\frac{T}{T_{0}}\bigg)\bigg].\label{eq:tandi}
\end{eqnarray}
Here $\delta f_0(T)=f_{0}-f(T)$, $\Psi$ is the di-gamma
function, $F$ is a geometric filling factor and $T_{0}$ is a reference
temperature. The measured data and fits to Eq. (\ref{eq:tandi}) are
shown in Fig. \ref{fig:tand}a. For all resonators we find $F\tan\delta_{i}=0.7\pm0.1\cdot10^{-5}$,
with no clear frequency dependence or variation between cool-downs
(verified in three cool-downs). The fitted loss tangents are plotted
in Fig. \ref{fig:tand}b together with the expected scaling for a
constant DOS ($\mu=0$) and a DOS with TLS interactions ($\mu=0.2$).
The scatter of the data is within the expected weak dependence of
the DOS on resonator frequency.

\clearpage
\section{$1/f$ frequency noise}
\begin{figure}[h!]
\includegraphics[scale=0.35]{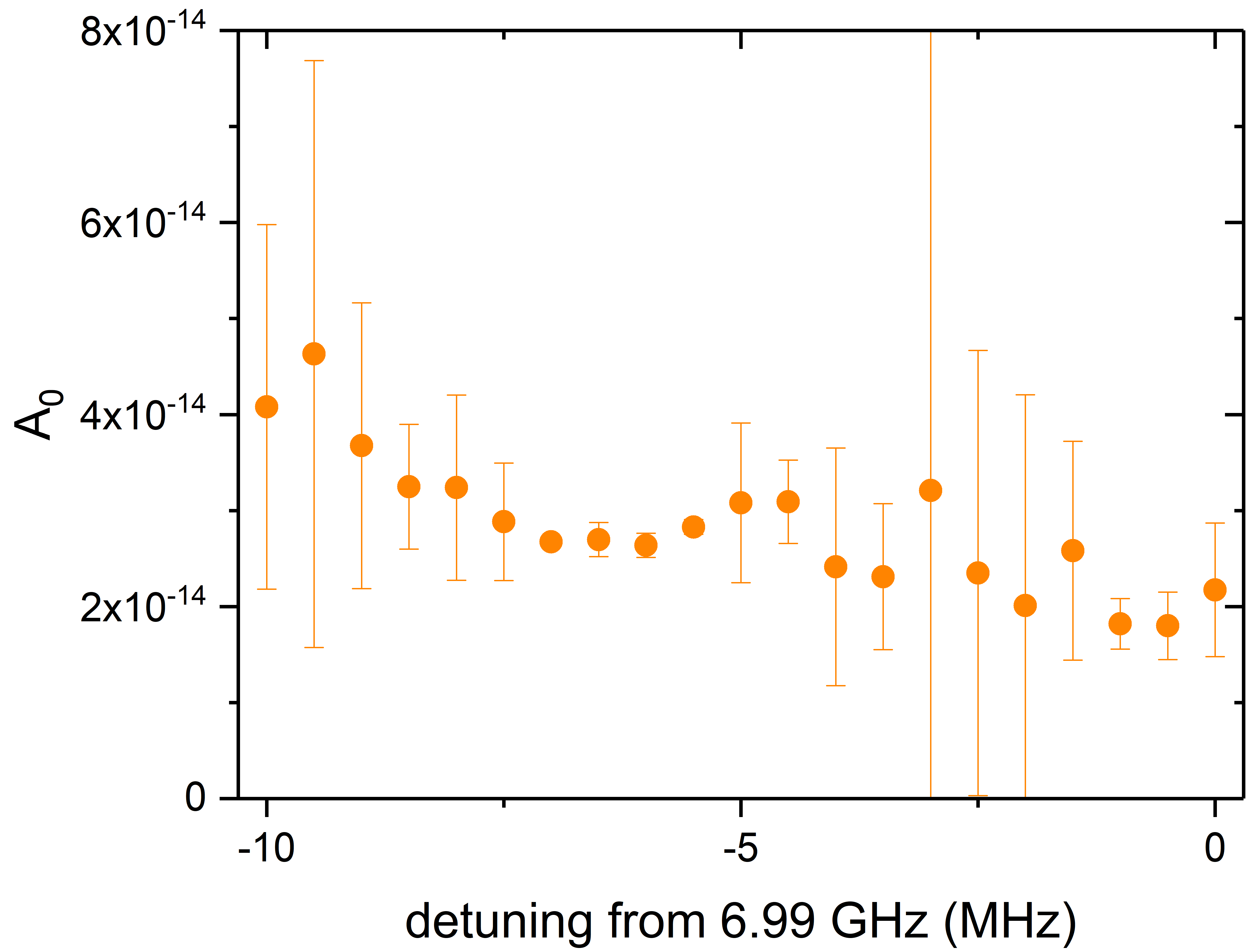} \caption{The magnitude $A_{0}$ of the $1/f$ frequency noise obtained from
the power dependence of the measured noise spectral density. Error
bars are the standard deviation of the calculated Allan Deviation
(ADEV) evaluated across timescales spanning two decades and propagated
into the fit to eq. (\ref{eq:sya}). A small error means that the
data is perfectly described by a $1/f$ power law whereas a large
error indicates a different noise process coming into play, e.g. a
Lorentzian fluctuator. All data obtained at 15 mK. }
\label{fig:freqnoise} 
\end{figure}

We also measure the frequency noise of the resonator to ensure that
the tuning capability does not induce additional frequency noise.
We do this using our Pound frequency locked loop setup described in
detail elsewhere \cite{tobiaspound,degraaf2018}. We observe clear
$1/f$ frequency noise for the KI tunable resonators of similar magnitude
as observed for other resonators of similar capacitor size \cite{burnett2016,degraaf2018,degraafAPL2018}.
We measure the frequency noise for a range of powers down to the single
photon regime and extract the magnitude of the TLS induced $1/f$
frequency noise, $A_{0}$, by fitting the individual measurements
in power to \cite{burnett2016} 
\begin{equation}
S_{y}(f)=\frac{A_{0}}{f\sqrt{1+\langle N\rangle/N_{c}}}.\label{eq:sya}
\end{equation}
Again we find similar values for $A_{0}$ as in previous studies.
We then repeat this measurement as a function of resonator frequency
detuning and plot $A_{0}(\delta f_{0})$ in Fig. \ref{fig:freqnoise}.
Indeed, the magnitude of the noise remains unaffected once the resonator is detuned from its base frequency, indicating that the kinetic inductance tuning does not affect the
resonator properties in any noticeable way. I.e. no additional kinetic indictance noise nor current noise in the biasing circuitry is large enough to dominate over the intrinsic TLS-induced $1/f$ noise. 
The larger error bars found for certain detunings are due to strongly coupled
TLS causing switching events and the noise deviates from the $1/f$-dependence
of the background TLS, and the extraction of the $1/f$ noise level becomes challenging.

\clearpage
\section{qTLS density in the presence of an electrostatic field}
\begin{figure}[h!]
\includegraphics[scale=0.6]{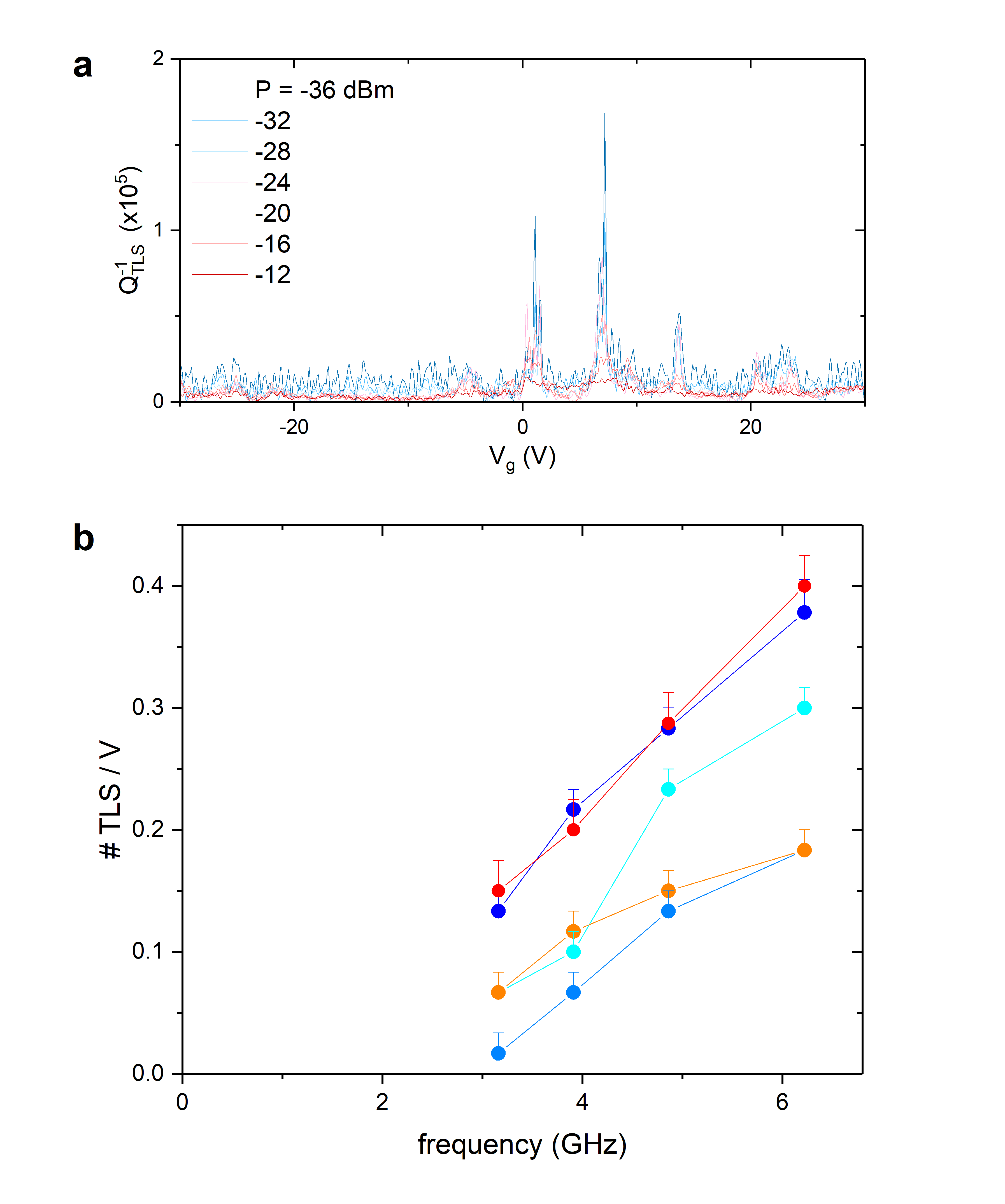} \caption{a) Measured loss rate of individual qTLS brought into resonance with the resonator by an applied electric field ($V_{g}$). Each trace corresponds to a different measurement power, in increments of 4 dB. Here $f_{0}=3.16GHz$.  The peaks are due to individual TLS coming into resonance with the resonator. b) Number of strongly coupled TLS as a function of frequency. The data
points are obtained by counting the number of qTLS observed as a voltage
is applied to a gate electrode located above the sample, tuning the
TLS energy by electric field. Multiple data points for each resonator
corresponds to separate cool-downs (different color). All data obtained
at 15 mK and $\langle N\rangle<100$. Variations in numbers between
cool-downs is most likely due to the random distribution of qTLS and
the limited range of fields explored. Error bars indicate miscounting the number of qTLS by one, e.g. one is situated just ourside the measurement window.}
\label{fig:densitygate} 
\end{figure}

We have also used a complementary technique to extract the qTLS density, by instead tuning the qTLS energies by an electrostatic gate. A gate voltage $V_{g}$ is applied to an electrode located in the
sample enclosure, and separated about $z_0 = 2.6$ mm from the surface of the
sample. The sample ground planes and resonator structures, PCB on
which it is mounted, and the sample enclosure form the counter electrode
(ground), similar to the setup in \cite{bilmes2019}. The electric field $E \sim V_g/z_0$ generated
at the sample surface couple to the TLS through their dipole moment
$p_0=qd_0$ and its projection angle $\theta$ along the direction of the
electric field, such that the TLS energy can be effectively described
by $E_{{\rm TLS}}=\sqrt{\epsilon_0^{2}+\gamma^2 V_{g}^{2}}$, where $\gamma = p_0\sin\theta /z_0$, 
tracing out a hyperbola as a function of applied electric field. The
static electric field distribution is expected to be the same for
all resonators. Fig. \ref{fig:densitygate}a shows a typical trace obtained when sweeping
the applied gate voltage, here with the background loss subtracted,
revealing several qTLS coming into resonance with the resonator. The data was obtained by keeping the resonator at zero
frequency detuning, and recording $Q_{i}$ as a function of $V_{g}$.
Fig. \ref{fig:densitygate}b shows the number of qTLS found for different
resonators when applying a gate voltage. 
Again, here we see a clear trend with much lower densities of qTLS
at low frequencies. Similarly to Fig. \ref{fig:linewidths}c in the main manuscript, we have here not
taken into account the additional scaling expected from the physical
footprint of the resonators, which would require an assumption about the exact locations of the qTLS.

\begin{figure}[t!]
\includegraphics[scale=0.35]{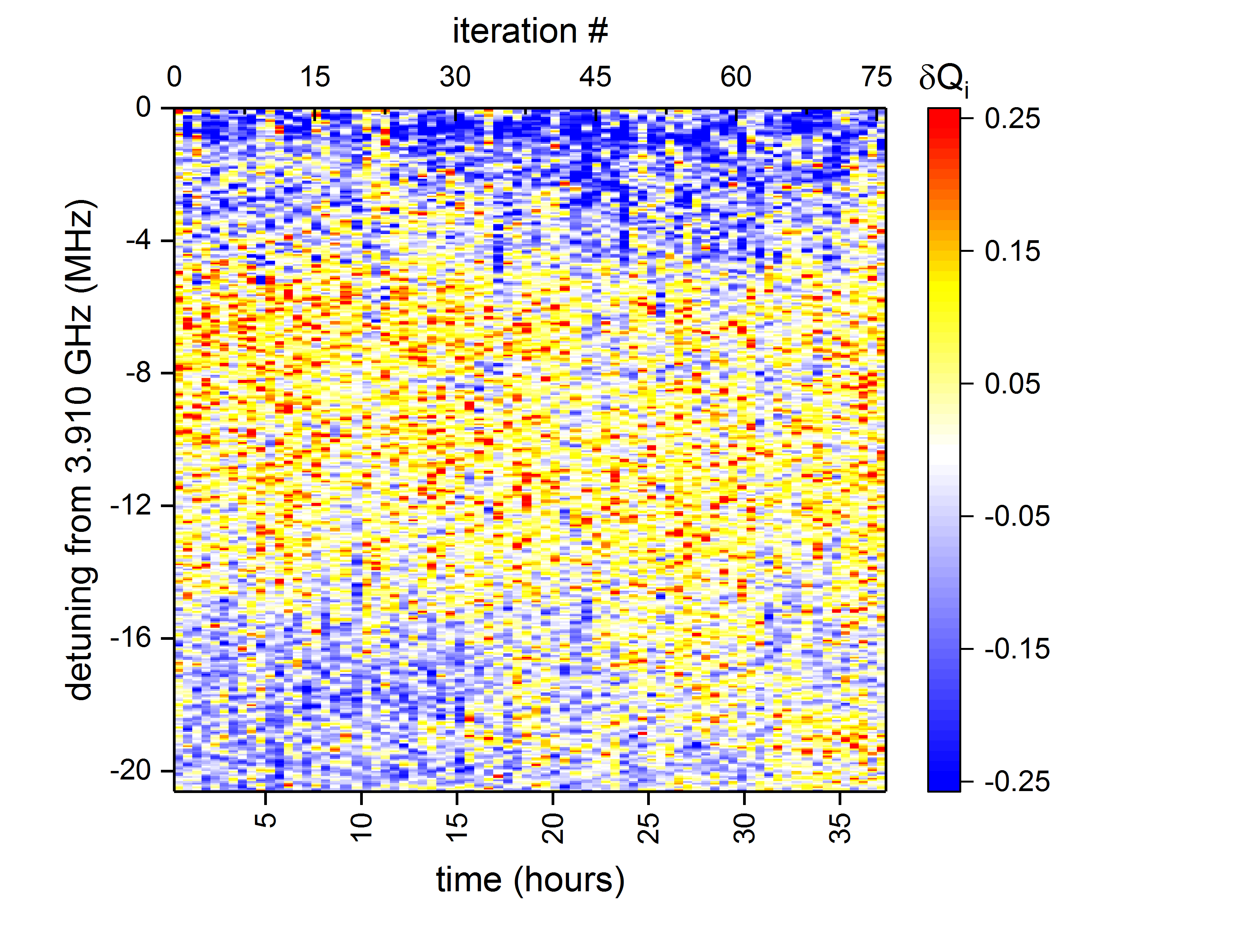} \caption{Spectral \& temporal fluctuations of the internal quality factor at
10 mK of a low frequency resonator. $\langle n\rangle\approx10$.
The color scale shows the variation in $Q_{i}$: $\delta Q_{i}=(Q_{i}-\langle Q_{i}\rangle)/\langle Q_{i}\rangle$,
where the average is taken across the whole measured parameter space.
The detuning is evaluated as the average fitted centre frequency across
all times.}
\label{fig:lowfspectraltemporal} 
\end{figure}

In Fig. \ref{fig:lowfspectraltemporal} we also show an example of
a spectral map from a 3.9 GHz resonator. It is clear that at this
frequency the number of TLS observed is much smaller, as compared to e.g. Fig. \ref{fig:spectraltemporalmultitemp} below. 

\clearpage
\section{Saturation of individual qTLS}
\begin{figure}[h!]
\includegraphics[scale=0.28]{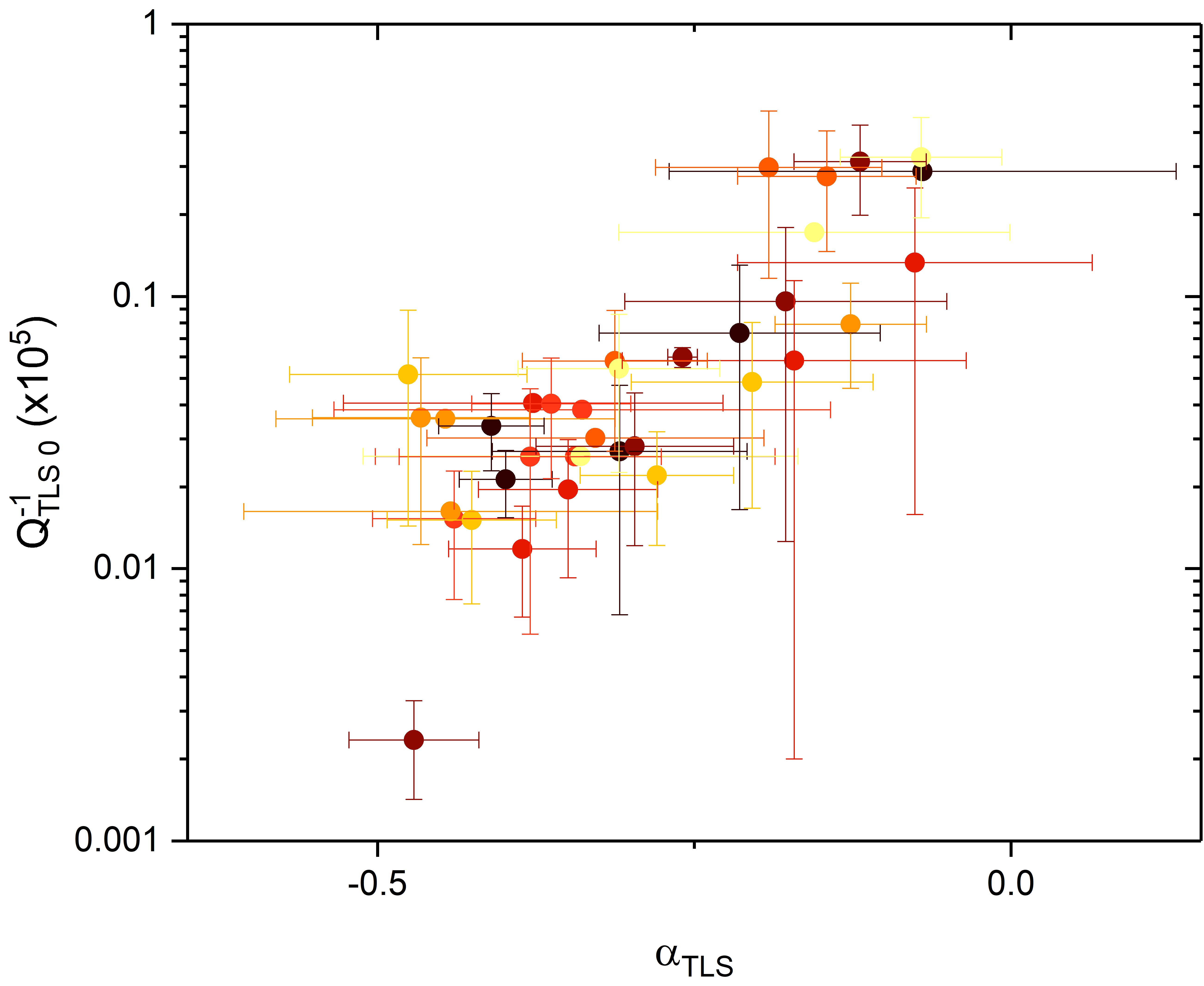} \caption{ 
Loss rate of individual TLS
observed through gate tuning, plotted against the power dependence of the TLS. Colors
correspond to different resonance frequencies: 6.22 GHz (yellow) through
3.16 GHz (black). Error bars are 95\% confidence bounds from the fit
to $Q_{{\rm TLS}}^{-1}(P)=Q_{{\rm TLS,0}}^{-1}P^{{\displaystyle \alpha_{{\rm TLS}}}}$.
All measurements done at 15 mK.}
\label{fig:tlssat} 
\end{figure}

We next show the saturation properties which further evidences the TLS nature of the observed fluctuations. Fig. \ref{fig:tlssat} shows the extracted power dependence for saturating
individually observed qTLS. For each qTLS we detect as a function of
applied gate voltage, we extract its power dependence. 
From the measured $Q_{i}(V_{g},P)$ we extract for each power the
loss rate induced by each qTLS: $Q_{{\rm TLS}}^{-1}=Q_{i}^{-1}(V_{g})-Q_{i,{\rm max}}^{-1}(V_{g})$.
We then fit the $Q_{{\rm TLS}}^{-1}(P)$ data to $Q_{{\rm TLS}}^{-1}(P)=Q_{{\rm TLS,0}}^{-1}P^{{\displaystyle \alpha_{{\rm TLS}}}}$
and plot the loss rate versus the power saturation exponent $\alpha_{{\rm TLS}}$
in Fig. \ref{fig:tlssat}.
The correlation between $\alpha_{{\rm TLS}}$ and $Q_{{\rm TLS}}$
is expected from the fact that qTLS with large dipole moments strongly
affect the resonator and, at the same time also interact strongly
with slow fluctuators. I.e. the more dissipative the TLS, the smaller
is $\alpha_{{\rm TLS}}$. 

\begin{figure}[h!]
\includegraphics[scale=0.26]{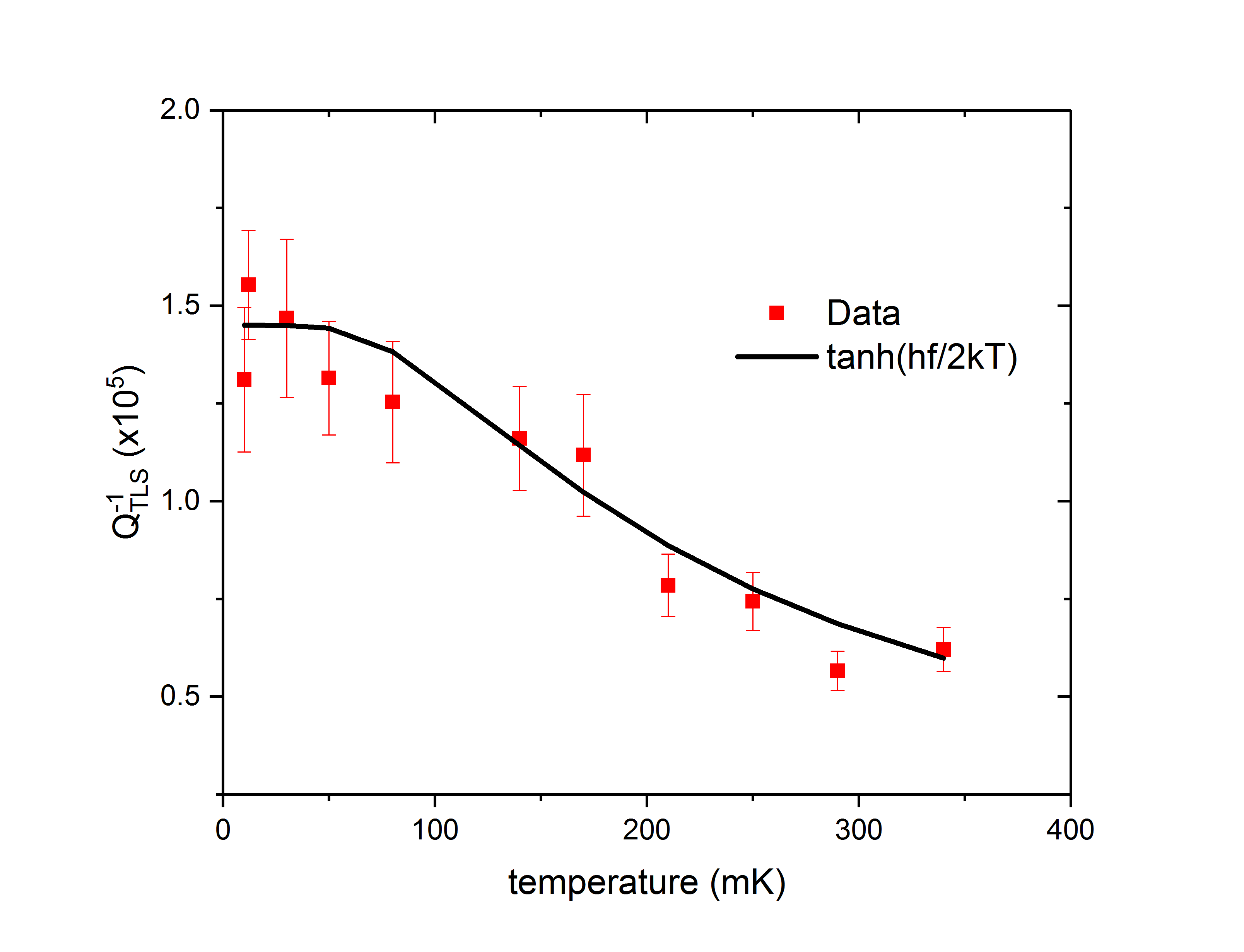} \caption{Temperature dependence of a single qTLS. Each datapoint is obtained
by tuning the qTLS through the resonance using an applied gate voltage.
The resulting peak in dissipation is then fitted to a Lorentzian with
the error bars indicating the 95\% confidence bounds. $\langle n\rangle\approx10$.}
\label{fig:tlstdep} 
\end{figure}

Fig. \ref{fig:tlstdep} also shows the measured temperature dependence
of the (low power, unsaturated) loss rate of a single qTLS tuned into
resonance using the electrostatic gate. The data fits well to
$Q_{{\rm TLS}}^{-1}\propto\tanh{(hf_{0}/2k_{B}T)}$ (we ignore the low power saturation regime), which is the expected temperature dependence of a single two-level system.

\clearpage
\section{Electron spin resonance spectrum}
\begin{figure}[h!]
\includegraphics[scale=0.3]{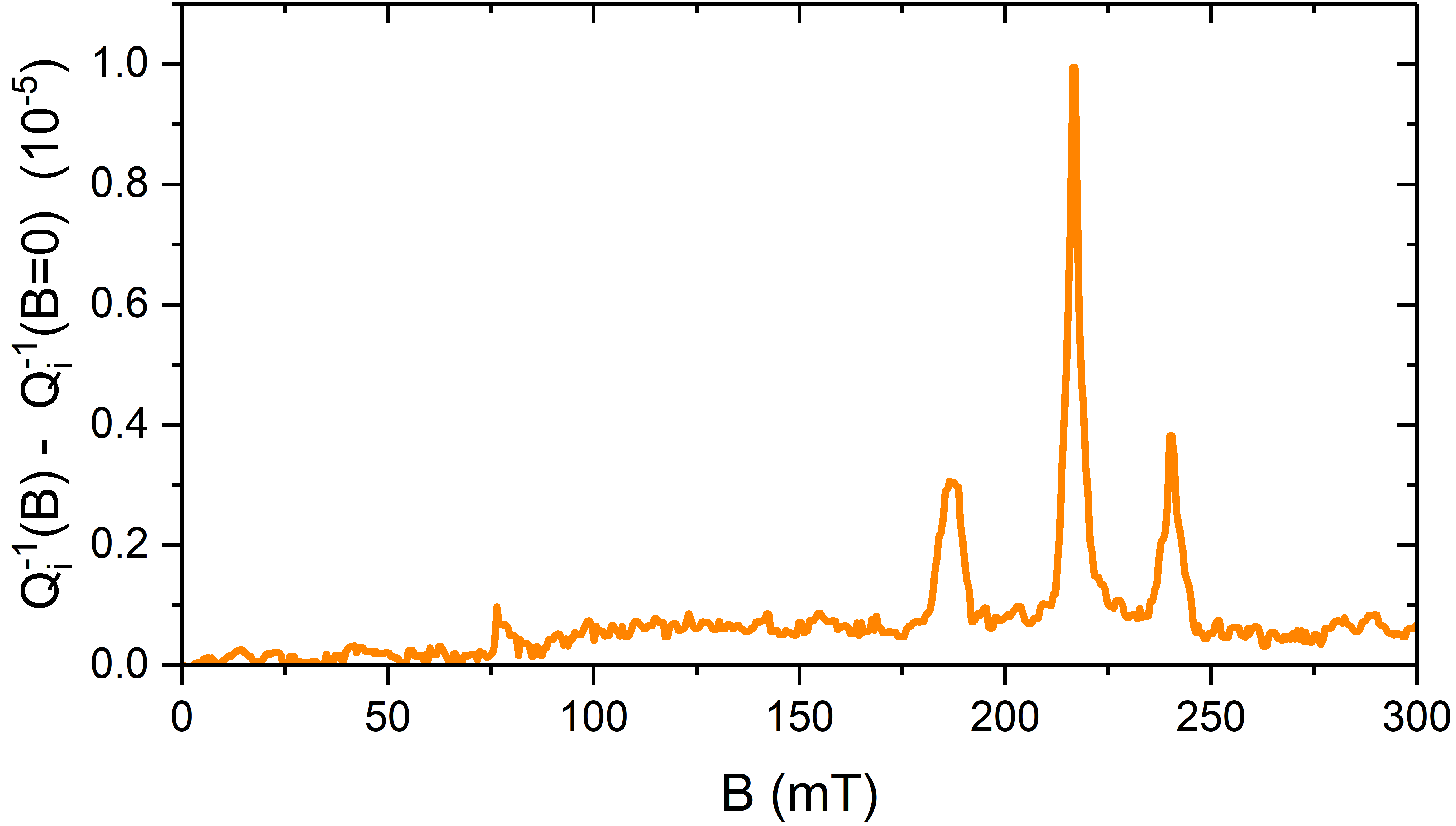} \caption{Intrinsic device on-chip ESR spectrum taken at 10 mK for a 5.87 GHz resonator, showing several features due to surface spins. The plotted quantity is the internal loss rate with the zero field loss rate subtracted.}
\label{fig:ESR} 
\end{figure}

Here we show using our constant wave on-chip electron spin resonance (ESR) technique that the samples
used here hold the same density and types of surface spins previously
analysed in detail \cite{degraaf2017}. This amount of surface spins is a likely source of the order parameter fluctuations. The ESR spectrum is shown
in fig \ref{fig:ESR}, where we plot the magnetic field induced loss
$1/Q_{i}(B)-1/Q_{i}(B=0)$. While some features can be attributed
to certain species \cite{degraaf2017}, there is still uncertanity in the broad background
plateau starting at about 100 mT, and the nature of the central 'g=2'
peak could, partially, be due to unpaired electrons on the device
surface, some of which are likely to be located on the surface of the superconductor. Further measurements will be required to elude any possible
ESR signature associated with the qTLS. 

\clearpage
\section{Thermal reconfiguration of qTLS}
\begin{figure*}[h!]
\begin{centering}
\includegraphics[scale=0.7]{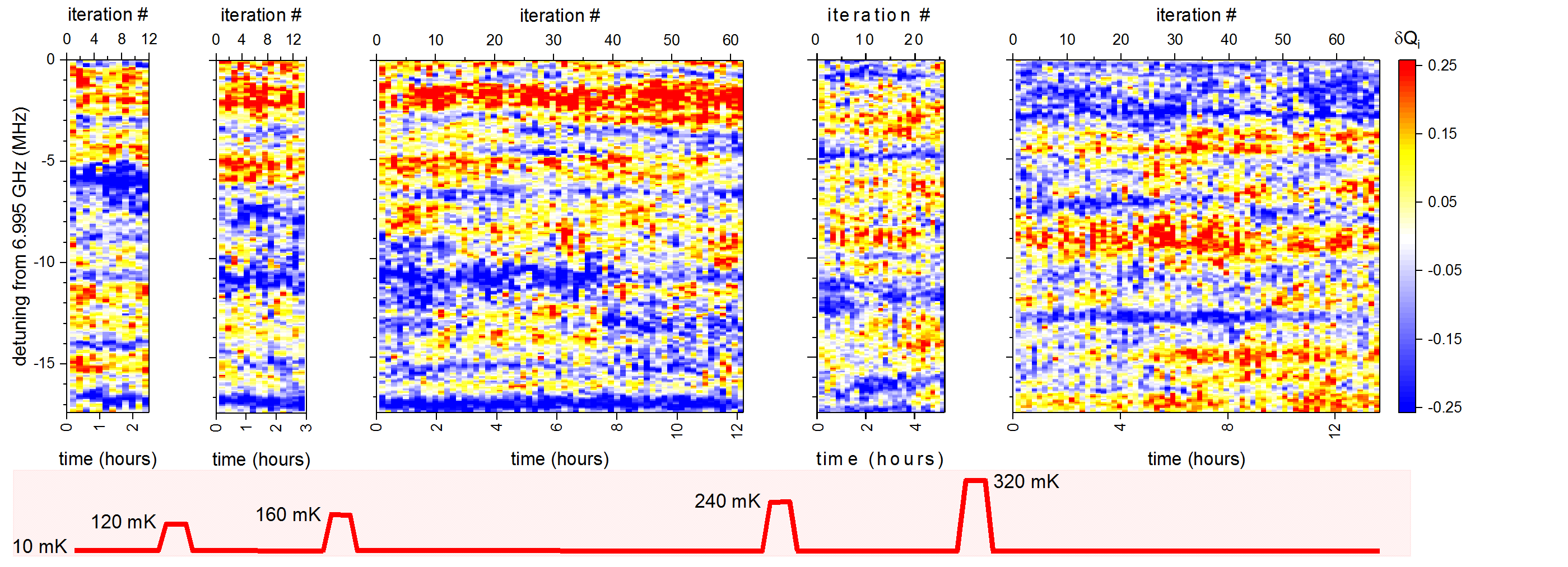}\hspace{5mm} \includegraphics[scale=0.64]{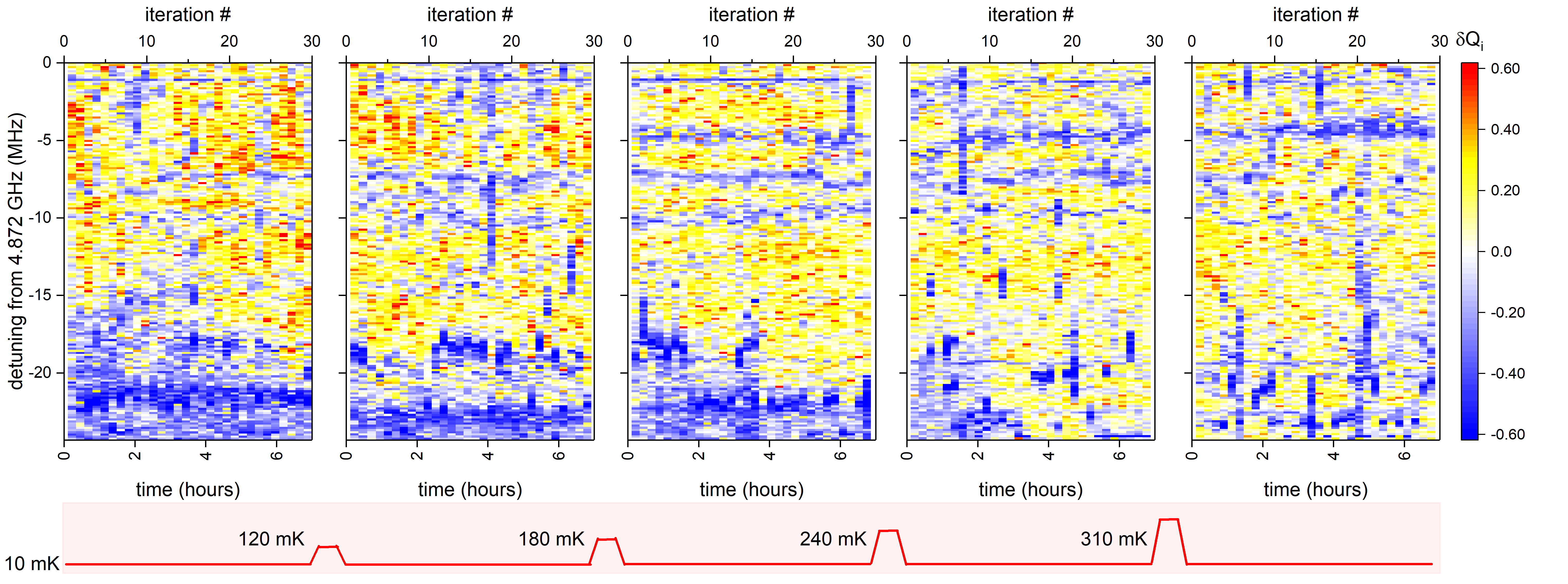} 
\par\end{centering}
\caption{Thermal reconfiguration of TLS. Spectral \& temporal fluctuations
of the internal quality factor at 10 mK after consecutive visits to
elevated temperatures for two different samples in two different cooldowns.
The data in the upper panels is from the same sample but a different
cool-down as that in Fig. \ref{fig:spectraltemporal} in the main
manuscript. All measurements taken at the same input power to the
resonator, corresponding to $\langle n\rangle\approx10$. The color
scale shows the variation in $Q_{i}$: $\delta Q_{i}=(Q_{i}-\langle Q_{i}\rangle)/\langle Q_{i}\rangle$,
where the average is taken across the whole measured parameter space.
The detuning is evaluated as the average fitted centre frequency across
all times.}
\label{fig:spectraltemporalmultitemp} 
\end{figure*}

\begin{figure*}[t!]
\includegraphics[scale=0.68]{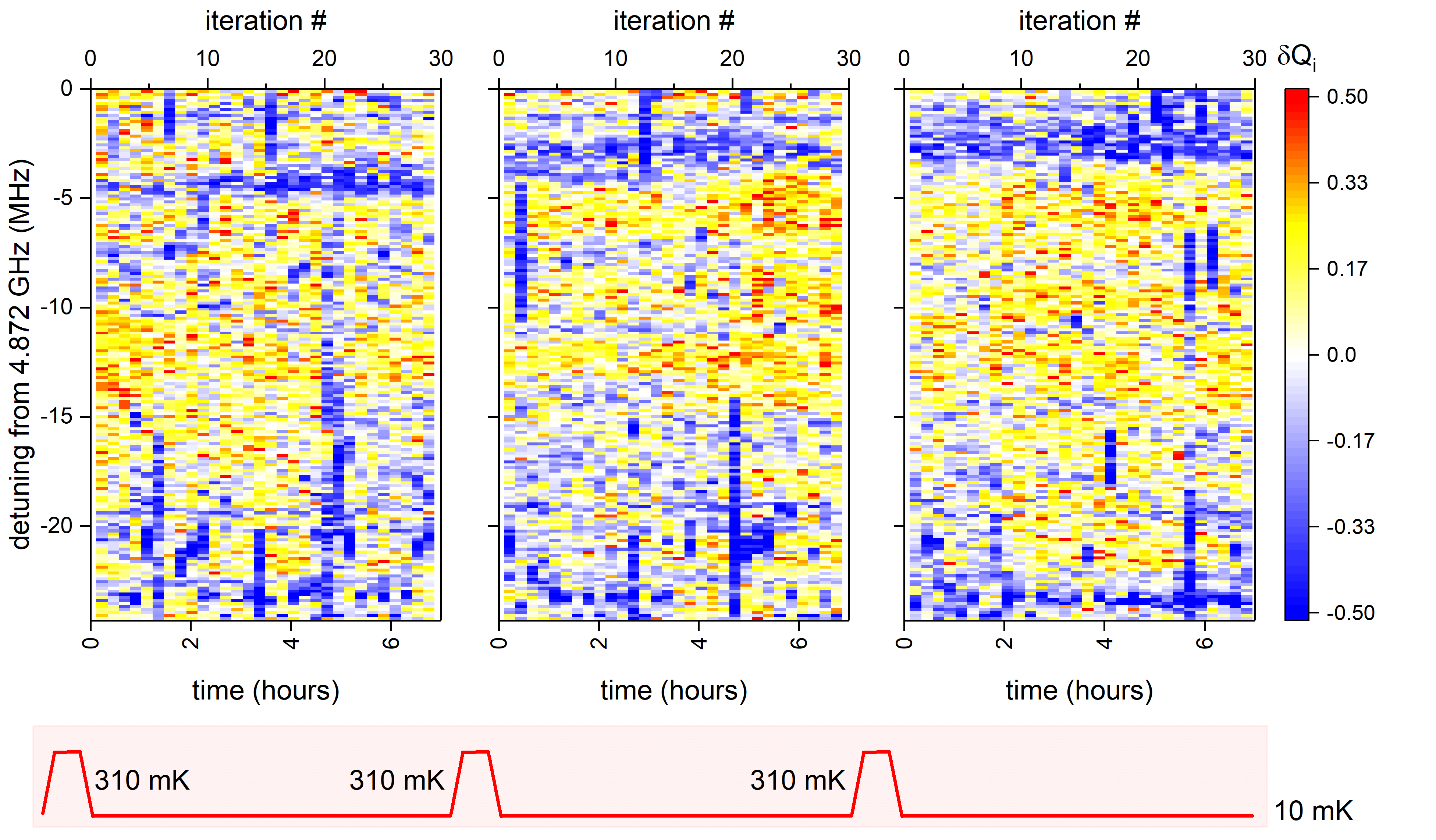} \caption{Thermal reconfiguration of TLS. Spectral \& temporal fluctuations
of the internal quality factor at 10 mK after repeated visits to 310
mK. All measurements taken at the same input power to the resonator,
corresponding to $\langle n\rangle\approx10$. The color scale shows
the variation in $Q_{i}$: $\delta Q_{i}=(Q_{i}-\langle Q_{i}\rangle)/\langle Q_{i}\rangle$,
where the average is taken across the whole measured parameter space.
The detuning is evaluated as the average fitted centre frequency across
all times.}
\label{fig:spectraltemporalrepeatedthermalcycling} 
\end{figure*}

Here we show additional data from repeated measurements of thermal cycling of the sample, experiments that all indicate a typical energy scale $\sim 300$ mK. Fig. \ref{fig:spectraltemporalmultitemp}a shows the spectral and temporal
dependence of the internal Quality factor of the same resonator that
is shown in Fig. \ref{fig:spectraltemporal} in the main manuscript.
The data in Fig. \ref{fig:spectraltemporalmultitemp}a was acquired
in a separate cool-down from room temperature, hence a very different
TLS configuration. We first perform a measurement at 10 mK, we then
raise the temperature to 120 mK for several hours (while acquiring
another measurement, data not shown). Returning to 10 mK we record
the second set of data. We then repeat this visiting 160, 240, and
320 mK. Again we only see a significant change in the TLS landscape
after visiting the highest temperature, 320 mK. The same measurement was repeated for a nother resonator on another sample, and is shown in \ref{fig:spectraltemporalmultitemp}b. Here, following the gradual increase in temperatures visited we also consecutively visited the highest temperature multiple times, shown in Fig. \ref{fig:spectraltemporalrepeatedthermalcycling}. This shows moderate stability between thermal cyclings on the timescale of $\sim 24$ hours, after the initial thermal reconfiguration. This would suggest that some qTLS traps where the quasiparticle initially escaped have not been re-populated, but instead different traps are filled due to a non-vanishing flux of new quasiparticles, and these may again be affected by repeated cycling to elevated temperatures. New quasiparticles available for trapping could for instance be generated through high-energy cosmic events or by affecting quasiparticles trapped in different parts of the device.

\clearpage
\section{Effect of tuning current on qTLS}
\begin{figure}[h!]
\includegraphics[scale=0.55]{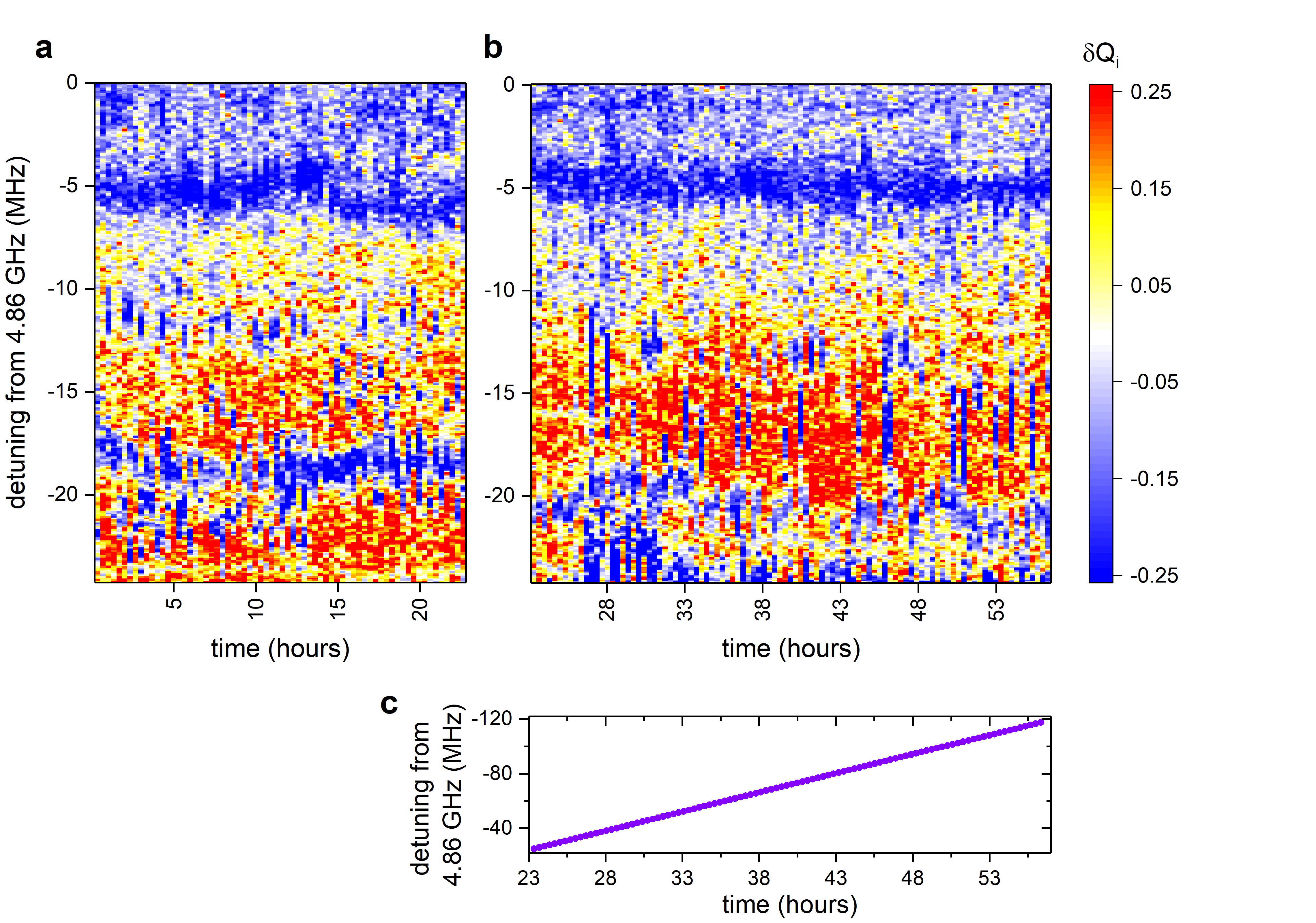} \caption{Effect of a large tuning current on the qTLS. a) Spectral and temporal
map of resonator quality factor fluctuations, similar to previous plots presented. The same measurement
is simply repeated over and over for 24 hours to verify the temporal
stability of observed qTLS. b) Just after the experiment in a) we
continue the spectral mapping, but after each measured trace the resonator
is detuned for 10 seconds to a higher frequency, as indicated in c),
up to near the critical current of the resonator.}
\label{fig:endpoint} 
\end{figure}

Another aspect to consider is whether the applied tuning current has any impact
on the observed qTLS. 
In Fig. \ref{fig:endpoint}a we first verify the temporal stability
of a few observed qTLS, similar to previous temporal plots shown. After
this experiment we immediately start another one (Fig. \ref{fig:endpoint}b)
using the same parameters, except after each frequency sweep of the
resonator we detune the resonator to a larger and larger frequency (for about 10 seconds),
the tuning current approaching the critical current of the superconducting
film. Naively one would expect the tuning current to have a similar effect as suppressing the gap, modifying locally the quasiparticle traps. The frequency to which the resonator was detuned is indicated
in Fig. \ref{fig:endpoint}c. Notably, we do not observe any major effect on the observed qTLS which could not be attributed to temporal instabilities in this experiment.
This is, however, to be expected since superconductivity
is only suppressed in the current-carrying parts of the resonator,
whereas the qTLS that can be detected are situated in the regions
of the resonator with large microwave electric fields, separated large distances from
the regions where the DC tuning current flows and the qTLS traps are already occupied. Thus, the shape of the qTLS potentials
that couple electrically is never altered, unless the critical current is exceeded and the resulting Joule heating leads to thermal reconfiguration.

\ifx\killheadings\undefined 

\clearpage
\section{Surface TLS density}
We now turn to the evaluation of the E-field numerical simulations of our device for the evaluation of the conventional background TLS density using several different approaches. 

First, following the method in \cite{burnett2016} we evaluate the density from the measured loss tangent. 
Assuming a dipole moment $p_{0}=0.8$ $\rm e\AA$ \cite{lisenfeld2019} and a
dielectric constant $\varepsilon_{r}=10$ of the TLS hosting medium
assumed to be within a surface layer of thickness 10 nm across the
whole resonator modal area we find the total number of conventional resonant TLS
coupling to the resonator $N_{{\rm TLS}}=\chi\Gamma_{2}V_{r}/U_{0}\approx430$,
where $V_{r}$ denotes the volume occupied by the TLS, $U_{0}=p_{0}^{2}/\varepsilon_{0}\varepsilon_{r}=10^{-2}$
eVnm$^{3}$, $\chi=F\tan\delta_{i}$ and $\Gamma_{2}\approx20\chi(k_{B}T)^{1+\mu}/(hf_{0})^{\mu}\approx7$
MHz at $T=50$ mK, where we have used the Filling factor $F=0.015$
and $\mu=0.2$ \cite{burnett2014}. This translates to $\rho_{{\rm TLS}}\approx60$
GHz$^{-1}$ ${\rm \mu}$m$^{-2}$, two orders of magnitude larger
than the number of observed individual strongly coupled qTLS. This is similar to what is typically found for this type of resonator \cite{burnett2016, degraaf2018}.

To evaluate the TLS density and contribution to device loss the above and other approaches
used so far are almost exclusively reliant on the evaluation of either
a participation ratio or filling factor of the dielectric volume of
interest. These approaches have two critical disadvantages when delaing
with very thin (and often unknown) layers of TLS on surfaces in the
form of thin native oxide layers or surface adsorbants. First, an
assumption about the layer thickness has to be made, and chosing the
correct thickness is not trivial, however, it can have a significant
impact on the final result. Furthermore, the dielectric constant of
the volume where the TLS are situated is required, and in particular
for a monolayer of surface adsorbates this is not very well defined.

Here we instead derive a direct approach to obtaining the TLS density
assuming all TLS of relevance reside on a surface. The only quantity
that needs to be determined is the electric field strength generated
by the device, at the surface of interest, and an assumption of the
density of TLS (here assumed to be constant both in energy and spatially).
This approach is valid for a distribution of TLS out of the plane
of the surface (i.e. thin layers of finite thickness), as long as
the electric field strength does not change significantly when the
TLS is located outside the evaluated surface. 

In the low power limit we can write the total TLS-limited loss as
the sum of loss originating from a larger number of TLS. 
\[
\frac{1}{Q_{{\rm TLS,tot}}}=\sum_{n}\frac{1}{Q_{{\rm n}}}.
\]
The absorption of a single TLS is given by its polarisation 
\[
{im}\langle S_{1}^{+}\rangle=\frac{\frac{1}{2}\tanh{\frac{\omega}{2k_{B}T}}\Omega\Gamma_{2}}{(\omega-\varepsilon)^{2}+\Gamma_{2}^{2}+\Omega^{2}\Gamma_{2}/\Gamma_{1}},
\]
where $\Omega={\frac{1}{2}{\bf p\cdot E}}/h$ is the Rabi frequency
due to the coupling of the TLS dipole to the resonator RF electric
field. In the low temperature, low drive regime we thus have 
\[
\frac{1}{Q_{{\rm TLS,tot}}}=\sum_{n}\frac{{\frac{1}{2}{\bf p_{n}\cdot E_{n}}}\Gamma_{2,n}/h}{\Delta_{n}^{2}+\Gamma_{2,n}^{2}}.
\]

First, integrating over all possible detunings (assuming a uniform
distribution in frequency of TLS) we get 
\[
\frac{1}{Q_{{\rm TLS,tot}}}=\frac{\pi}{2h}\rho_{\Delta}\sum_{n}{\bf p_{n}\cdot E_{n}},
\]
where $\rho_{\Delta}$ is the energy density of TLS (in number of
TLS per Hz). We then assume the ensemble average $p_{0}$ of dipole
moments 
\[
\frac{1}{Q_{{\rm TLS,tot}}}=\frac{\pi p_{0}}{2h}\rho_{\Delta}\sum_{n}{|{\bf E}_{n}||\sin\theta_{n}|},
\]
where $\theta_{n}$ is the angle between each TLS dipole and the local
electric field. As the RF electric field is perpendicular to the metal
surface, the averaging of $\theta$ goes over all possible angles
of the TLS orientation only (an approximation that holds for the metal
surface but not the substrate-air). The local electric field can be
expressed as ${\bf E}=E(x,y)$. We can then convert the sum to an
integral over the surface of the superconductor by assuming a uniform
surface density (combined with the energy density) $\rho_{{\rm \Delta,A}}$
of TLS 
\[
\frac{1}{Q_{{\rm TLS,tot}}}=\frac{\pi p_{0}}{2h}\rho_{{\rm \Delta,A}}\int_{A}\int_{0}^{2\pi}E(x,y)|\sin\theta|dxdyd\theta.
\]
We can decompose
the electric field into a component running along the resonator and
one component along the perimeter of the metal. Along the length $L$
of the resonator the field is decaying sinusodially ($\lambda/2$
or $3\lambda/2$ resonator), such that the integrand in $y$ becomes
$L/2$.

\[
\frac{1}{Q_{{\rm TLS,tot}}}=\frac{\pi p_{0}}{2\sqrt{2}h}\rho_{{\rm \Delta,A}}\frac{L}{2}\int E(x)dx.
\]
The remaining integral is carried out in COMSOL by evaluating the
electric field around the surface of the superconductor. Details of such simulation for our device is presented in Fig. \ref{fig:RFfieldstrength}. For a single
photon maximum RF field amplitude of $V_{ph}=\sqrt{4\pi hf_{0}^{2}Z}=3.5$
${\rm \mu}$V the integral takes in our case the value $\int E(x)dx=2\cdot10^{-5}$
V around the perimeter of the superconductor (cross-section), including
both the M-A and S-M interfaces. Assuming $p_{0}=0.8$ e\AA, we get
for $f_{0}=4.9$ GHz 
\[
\frac{1}{Q_{{\rm TLS,tot}}}=0.003\rho_{{\rm \Delta,A}}.
\]
For the measured single photon $Q_{{\rm TLS,tot}}=10^{5}$ we get
$\rho_{{\rm \Delta}}\approx3.4\cdot10^{6}$ GHz$^{-1}$ integrated
over the area of the resonator, or $\sim330$ resonant TLS (within
the bandwidth of the resonator). Dividing out the area of the surface
of the superconductor in our resonator we find $\rho_{{\rm \Delta,A}}\approx50$
GHz$^{-1}$ ${\rm \mu}$m$^{-2}$. These resonant TLS are weakly coupled
and each TLS do not contribute significantly to the total loss, for
instance these could be the vast majority of TLS with a dipole moment
mostly perpendicular to the electric field oriantation.

In the high power limit we instead get 
\[
\frac{1}{Q_{{\rm TLS,tot}}}=\frac{\pi}{2}\rho_{\Delta}\sum_{n}\frac{\Omega_{n}}{\sqrt{1+\Omega_{n}^{2}/\Gamma_{1,n}\Gamma_{2,n}}},
\]
where we can identify a 'critical' field strength for saturation of
the TLS $E_{n,crit}=h\sqrt{\Gamma_{1,n}\Gamma_{2,n}}/p_{n}$. 

\begin{figure*}[t!]
\includegraphics[scale=0.6]{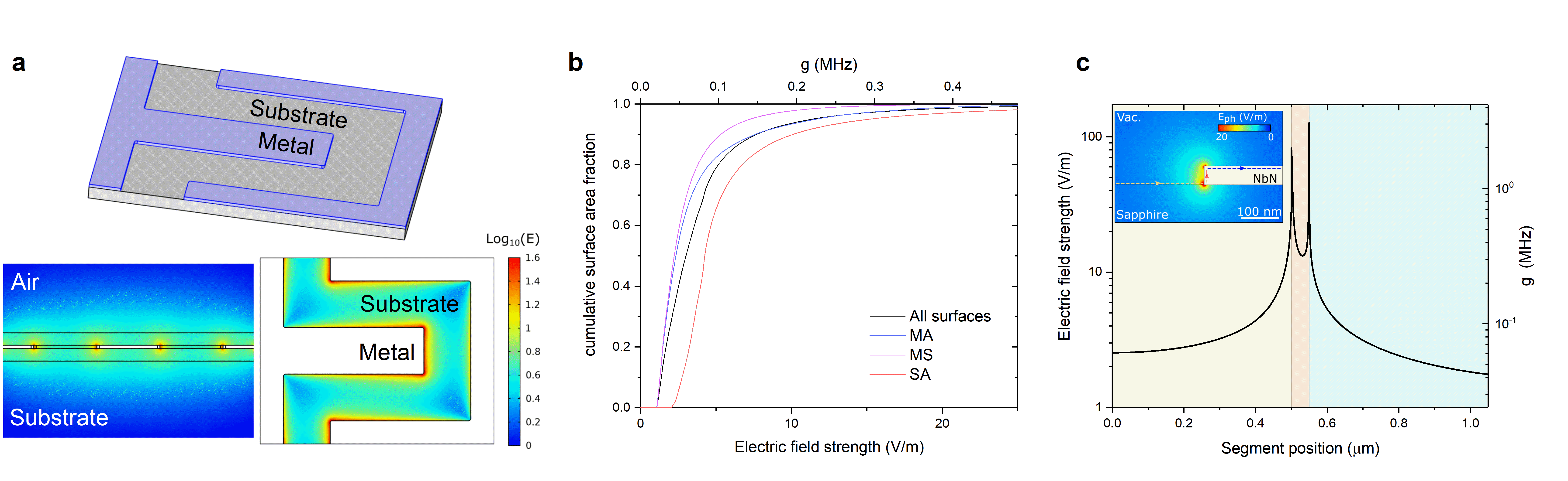} \caption{Microwave electric field strength in the resonator. a) Single photon electric field strengths obtained through a COMSOL simulation
for a segment of the resonator, assumed to be at the voltage antinode of the resonator. Top: 3D model of the simulated volume. Bottom left: vertical cross-section showing Electric field strengths. Bottom right: Cross section in the substrate surface plane. b) integrated probability distributions for elenctric field strengths and corresponding resonator-TLS coupling strengths (assuming dipole moment $p_0 = 0.8$ e\AA) for the different device interfaces; Metal-Substrate (MS), Metal-Air (MA) and Substrate-Air (SA). c) Cross-sectional evaluation of the electric field strength near a metal edge showing that significant coupling strengths ($>50$ kHz $\gtrsim f_0/Q$) can be obtained across the whole metal surface of the device.}
\label{fig:RFfieldstrength} 
\end{figure*}

\ifx\killheadings\undefined 

\clearpage
\section{qTLS magnetic field measurements}
Finally, we here outline the details of the qTLS measurement in mangetic field shown in Fig. 4 of the main manuscript. To measure a coherent qTLS in magnetic field we placed the sample
in the bore of a 9/1/1 T vector magnet inside the dilution refrigerator.
The resonators themselves are highly magnetic field resilient, retaining
quality factors in excess of $10^{5}$ above $B=1$ T. To ensure optimal
alignment of the magnetic field with the superconducting plane we
applied a small field (5 mT), by rotating the field by a small amount
(few degrees) and tracking the resonance frequency as a function of
field angle we find the optimal orientation from the maximum in the
resonance frequency.

The experiments presented in the main manuscript were carried out
at zero current detuning, and we instead tune the TLS into resonance
with the resonator using the electrostatic gate in the sample enclosure.
The reason for this was purely technical, for larger tuning currents
the resonator becomes more sensitive to current noise, and most likely
due to mechanical resonances and eddy currents in the coaxial cables
excited by the pulse tubes of the cryostat, we saw an increased instability
of the resonator frequency at larger magnetic fields. By staying at
zero current detuning we minimise this noise.

\end{document}